\documentclass[preprint,onecolumn,aps,nofootinbib]{revtex4}
\usepackage[T1]{fontenc}
\usepackage{rotating}
\usepackage{multirow}
\usepackage{lscape}
\usepackage{appendix}
\usepackage{subfigure}
\usepackage{graphicx}
\usepackage[cmyk]{xcolor}
\usepackage{amsmath}
\usepackage{amsfonts}
\usepackage{amssymb}
\usepackage{MnSymbol}
\usepackage{wasysym}
\usepackage{braket}
\usepackage{diagbox}
\usepackage{eurosym}
\usepackage{calrsfs}
\usepackage{booktabs}
\usepackage{dcolumn}
\usepackage{ulem}
\usepackage[colorlinks=true,linkcolor=blue,urlcolor=blue,filecolor=black,citecolor=red,pdfstartview=FitV]{hyperref}
\newcommand{\red}[1]{{\color{red}{#1}}}

\begin{document}
\title{\boldmath Holographic striped superconductor with ionic lattice}

\author{Kai Li $^{1,2}$}
\email{lik@ihep.ac.cn}
\author{Yi Ling $^{1,2}$}
\email{lingy@ihep.ac.cn}
\author{Peng Liu $^{3}$}
\email{phylp@email.jnu.edu.cn}
\author{Meng-He Wu$^{4}$}
\email{mhwu@njtc.edu.cn} 

\affiliation{  
    $^1$ Institute of High Energy Physics, Chinese Academy of Sciences, Beijing 100049, China;\\
    $^2$ School of Physics, University of Chinese Academy of Sciences, Beijing 100049, China;\\
    $^3$ Department of Physics and Siyuan Laboratory, Jinan University, Guangzhou 510632, China;\\
     $^4$  College of Physics and Electronic Information Engineering, Neijiang Normal University, Neijiang 641112, China.
}

\begin{abstract}
We construct a holographic model to study the striped superconductor on ionic lattices. This model features a phase diagram with three distinct phases, namely the charge density wave (CDW) phase, ordinary superconducting phase (SC) and the striped superconducting phase (SSC). The effect of the ionic lattices on the phase diagram is investigated in detail. First, due to the periodic nature of the background, different types of CDW solutions can be found below the critical temperature. Furthermore, with the increase of the lattice amplitude these solutions are locked in different commensurate states. Second, we find that the critical temperature of CDW phase decreases with the increase of the lattice amplitude, while that of the SC phase increases. Additionally, the background solutions are obtained for different phases, and it is verified that the SSC phase has the lowest free energy among all three phases.

\end{abstract}
\maketitle
\tableofcontents

\section{Introduction}\label{sec:intro}
The striped superconductor as a novel superconducting phase plays a vital role in understanding the mechanism of high-temperature superconductivity. This phase is characterized by a unidirectional pair-density-wave (PDW) \cite{PhysRevB.79.064515,Berg_2009,2019arXiv190409687A,annurev:/content/journals/10.1146/annurev-conmatphys-031119-050711}, which emerges as the result of the interplay between the SC and CDW order \cite{2018NatCo...9.2796C}. Fundamentally, the CDW phase is an insulating phase reflecting the instability of the Fermi surface due to the interaction between electrons and lattices, while the SC phase is metallic with a divergent DC conductivity due to the formation of Cooper pairs.
Near the Fermi energy, these two orders often display complex relationships that can be competitive, cooperative, or both \cite{2024PhRvM...8a4801L}.
In particular, due to the involvement of strong coupling in the system, it is widely believed that such relations may not be fully understood within the traditional Landau-Fermi liquid theory \cite{Liu:2009dm}. The SSC phase as the product of the interplay of CDW order and SC order remains mysterious in the condensed matter theory \cite{PhysRevB.28.2855,osti_1357580}.

Gauge/gravity duality provides powerful tools for studying the striped superconductor since this strongly coupling many-body system living on the boundary can be dual to a weakly coupled gravity system in the bulk \cite{Hartnoll:2016apf}. Great efforts have been made to explore the novel properties of the striped superconductor using holographic approaches \cite{Flauger:2010tv,Ganguli:2012up,Donos:2013gda,Erdmenger:2013zaa,Donos:2013wia,Withers:2013loa,Ling:2014saa,Kiritsis:2015hoa,2017PhRvD..95d1901C,Cremonini:2017usb,Cai:2017qdz,Ling:2020qdd,Ling:2015epa,Baggioli:2015zoa,Ling:2015exa,Baggioli:2015dwa,Baggioli:2016oju,Cremonini:2016rbd,Andrade:2017cnc,Alberte:2017oqx,Amoretti:2017axe,Amoretti:2017frz,Cai:2020nyd}. 
However, early models are characterized by the emergence of both the CDW and SC phases with a single order parameter, implemented through the Stuckelberg mechanism. This approach implies
that both phases would emerge simultaneously below the same critical temperature.
This of course is not the case occurring in laboratory. Fundamentally, the CDW phase and SC phase should be independent and thus be described by different order parameters \cite{2024PhRvM...8a4801L}. In particular, from the symmetry perspective, the CDW phase arises from the spontaneous breaking of the translational symmetry while the SC phase results from the spontaneous breaking of $U(1)$ gauge symmetry. This distinction underscores the need for separate order parameters in accurate models. Until recently a holographic model with different order parameters for CDW phase and SC phase has been constructed in \cite{Ling:2020qdd}, where a phase diagram has been obtained for the interplay of these three phases. It demonstrates that  the critical temperature of SC increases with doping parameter while the critical temperature of CDW decreases, which mimics the behavior of some known materials \cite{PhysRevB.28.2855,osti_1357580}.

In this paper, we extend the previous work by investigating striped superconductors in the presence of an ionic lattice. The motivations of this work are twofold. First, the lattice is an essential structure for any practical material to generate momentum relaxation, and thus leading to a finite DC conductivity at finite temperature \cite{Horowitz:2012gs,Horowitz:2012ky,Horowitz:2013jaa,Ling:2013aya,Ling:2013nxa}. Thus, in contrast to the work in \cite{Ling:2020qdd}, we will investigate the spontaneous breaking of the translation symmetry as well as $U(1)$ gauge symmetry over an ionic lattice. 
In particular, we will focus on the impact of the lattice on the phase diagram and discuss its possible connection to the feature of real materials. Second, simulating Mott insulator in commensurate states is the key step to understanding the phase diagram of high temperature superconductivity. Recent progress in holographic duality reveals that it is essential to incorporate inhomogeneous lattices rather than homogeneous lattices into the bulk geometry to implement the commensurate lock-in effect \cite{Andrade:2015iyf,Andrade:2017leb,Andrade:2017jmt,Andrade:2017ghg,Krikun:2017cyw,Ling:2023sop}. Currently, the commensurate lock-in effect has been investigated only in several holographic models, where no superconducting phase is considered \cite{Andrade:2017leb,Ling:2023sop}.  Therefore, in this paper, we intend to build a striped superconductor on an ionic lattice, which could be viewed as the extension of our work \cite{Ling:2020qdd}, where the striped superconductivity is implemented but the lattice is absent. On the other hand, we intend to apply the ionic lattice as the typical inhomogeneous lattice to observe the commensurate lock-in effect in striped superconductor, which could be viewed as the extension of our work in \cite{Ling:2023sop}, where the superconducting phase is absent.

The paper is organized as follows: In Sec. \ref{sec2}, we introduce the holographic setup for the striped superconductor with ionic lattice and incorporate the concept of doping through the implementation of two U(1) gauge fields. In Sec. \ref{sec3}, we study the CDW phase over the ionic lattice by considering the instability of the background via linear perturbation. The phase diagram is obtained by evaluating the critical temperature under different values of the doping parameter. More importantly, we demonstrate two distinct types of CDW solutions  by numerically solving the equations of motion with full backreaction, and the feature of order parameter as well as the charge density is analyzed. In Sec. \ref{sec4}, we study the SC phase in parallel by considering the spontaneous breaking of $U(1)$ gauge symmetry separately. In Sec. \ref{sec5}, we investigate the SSC phase in detail by considering the interplay between the CDW order and the SC order. The background with full backreactions are obtained numerically and the behavior of the order parameters is demonstrated. We also justify that among all three phases  the SSC phase exhibits the best stability by computing the free energy. We present the conclusions and discussions in the last section.

\section{Holographic setup}\label{sec2}
In alignment with the work presented in \cite{Ling:2020qdd}, we consider a four-dimensional holographic model which contains the metric tensor $g_{\mu\nu}$, a real scalar (dilaton) field $\Phi$, two U(1) gauge fields $A_\mu$ and $B_\mu$, and a complex scalar field $\Psi$. The action for the system is given by:
\begin{equation}
    \begin{aligned}
        S= & \frac{1}{2 \kappa^2} \int d^4 x \sqrt{-g}\left[R-\frac{1}{2}(\nabla \Phi)^2-V(\Phi)-\frac{1}{4} Z_A(\Phi) F^2-\frac{1}{4} G^2\right. \\
           & \left.-|(\nabla-i e B) \Psi|^2-m_v^2 \Psi \Psi^*\right],
    \end{aligned}
\end{equation}
where $F = dA$ and $G = dB$ are the field strengths of the gauge fields $A_\mu$ and $B_\mu$, respectively. The coupling function $Z_A(\Phi)=1-\frac{\beta}{2} L^2 \Phi^2$ and the potential $V(\Phi)=-\frac{1}{L^2}+\frac{1}{2} m_s^2 \Phi^2$ are specified for the dilaton field $\Phi$, where $L$ is the AdS radius. Here, $m_s$ and $m_v$ are masses of the scalar field $\Phi$ and $\Psi$, respectively. In addition, $\kappa$ is the gravitational coupling constant, and $R$ is the Ricci scalar, and $g$ is the determinant of the metric tensor. In this model, the nonzero solution of $\Phi$ signals the spontaneous breaking of the translational symmetry and thus serves as the order parameter for the CDW phase. The presence of the term $\beta$ in $Z_A(\Phi)$ is to induce the instability of the background with vanishing $\Phi$. While $\Psi$ is a complex scalar field introduced to achieve spontaneous breaking of the U(1) gauge symmetry, and its norm is denoted as the order parameter of the SC phase. As usual, we redefine $\Psi$ as $\eta e^{i \theta}$, where $\eta > 0$ and the phase $\theta$ is fixed to zero by gauge choice. In line with the analysis in \cite{Ling:2020qdd,Ling:2023sop}, we treat the gauge field $B_\mu$ as the electromagnetic field dual to the boundary theory.

The equations of motion can be derived directly as follows, The equations of motion can be derived directly as follows, where prime ($\prime$) represents derivative with respect to $\Phi$,
\begin{equation}\label{eom1}
    \begin{aligned}
        R_{\mu \nu}-T_{\mu \nu}^{\Phi}-T_{\mu \nu}^A-T_{\mu \nu}^B-T_{\mu \nu}^\eta & =0, \\
        \nabla^2 \Phi-\frac{1}{4} Z_A^{\prime} F^2-V^{\prime}                       & =0, \\
        \nabla^2 \eta-m_v^2 \eta-(e B)^2 \eta                                       & =0, \\
        \nabla_\mu\left(Z_A F^{\mu \nu}\right)                                      & =0, \\
        \nabla_\mu G^{\mu \nu}-2 e^2 \eta^2 B^\nu                                   & =0.
    \end{aligned}
\end{equation}
where
\begin{equation}\label{eom}
    \begin{aligned}
        T_{\mu \nu}^{\Phi} & =\frac{1}{2} \nabla_\mu \Phi \nabla_\nu \Phi+\frac{1}{2} V g_{\mu \nu},                       \\
        T_{\mu \nu}^A      & =\frac{Z_A}{2}\left(F_{\mu \rho} F_\nu^\rho-\frac{1}{4} g_{\mu \nu} F^2\right),               \\
        T_{\mu \nu}^B      & =\frac{1}{2}\left(G_{\mu \rho} G_\nu^\rho-\frac{1}{4} g_{\mu \nu} G^2\right),                 \\
        T_{\mu \nu}^\eta   & =\nabla_\mu \eta \nabla_\nu \eta+e^2 \eta^2 B_\mu B_\nu+\frac{1}{2} m_v^2 \eta^2 g_{\mu \nu}.
    \end{aligned}
\end{equation}

Throughout this paper we will introduce the lattice structure and consider the translational symmetry breaking along $x$ direction only, thus we adopt the following ansatz for the metric, gauge fields, and scalar fields \cite{Horowitz:2012gs,Horowitz:2012ky,Horowitz:2013jaa,Ling:2013aya,Ling:2013nxa,Ling:2014saa,Ling:2020qdd,Ling:2023sop}:
\begin{equation}
    \begin{aligned}
        d s^2 & =\frac{1}{z^2}\left[-(1-z) p(z) Q_{t t} d t^2+\frac{Q_{z z} d z^2}{(1-z) p(z)}+Q_{x x}\left(d x+z^2 Q_{x z} d z\right)^2+Q_{y y} d y^2\right], \\
        A_t   & =\mu_1(1-z) a, \quad B_t=\mu_2(x)(1-z) b, \quad \Phi=z \phi, \quad \eta=z \zeta.
    \end{aligned}
\end{equation}
where $p(z)=4\left(1+z+z^2-\frac{\left(\mu_1^2+\mu_2^2\right) z^3}{16}\right), \mu_1$ and $\mu_2$ are the chemical potential of the gauge field $A$ and $B$, respectively. Here, $Q_{t t}, Q_{z z}, Q_{x x}, Q_{y y}, Q_{x z}, \psi, a, b, \phi, \zeta$ are functions of $x$ and $z$. In the special case where $Q_{tt}=Q_{zz}=Q_{xx}=Q_{yy}=a=b=1$ and $Q_{xz}=\phi=\zeta=0$, the metric reduces to a doubly charged version of the AdS-RN black hole. Throughout this paper, we set the AdS radius $l^2 = 6L^2 = 1/4$, the masses of the dilaton and the complex scalar field $m_s^2 = m_v^2 = -2/l^2 = -8$, and the coupling constant $\beta = -129$. Additionally, to ensure the large $N$ limit, we take $\frac{l^2}{2\kappa^2} \gg 1$. We adopt $\mu_1$ as the scaling unit and define $ X = \mu_2 / \mu_1 $ as the doping parameter. The Hawking temperature of the black hole is given by,
\begin{equation}
    \frac{T}{\mu_1} = \frac{48 - \mu_1^2 - \mu_2^2}{16 \pi \mu_1}.
\end{equation}

In the absence of lattices, the striped superconducting phase in this model has previously been investigated in \cite{Ling:2020qdd}. Here we intend to introduce an ionic lattice to explicitly break the translational symmetry of the background. Without loss of generality, we build an ionic lattice along $x$ direction by introducing a spatially modulated chemical potential for gauge field $B$ at $z\rightarrow 0$,
\begin{equation}
    \mu_2(x)=\mu_2+\mu_1 \lambda \cos (k x)=\mu_1\left(X+\lambda \cos( k x)\right),
\end{equation}
where $\lambda$ and $k$ are the amplitude and wave-vector of the ionic lattice, respectively.  In addition, we remark that since $\mu_1$ is treated as the scaling unit, we set the lattice amplitude with dimension as $\lambda\mu_1$ rather than $\lambda\mu_2$. As a result, the strength of the lattice is independent of the doping parameter $X$ in this setup, and one can adjust the lattice amplitude and the doping parameter separately.  Specially, the lattice structure can be maintained even if $X=0$.  Taking this as an input boundary condition, the solutions for the background exhibiting a periodic structure along $x$ direction can be obtained by numerically solving the equations of motion. Also, we will solve the Einstein equation in (\ref{eom1}) with Einstein-DeTurck method  \cite{2010CQGra..27c5002H,Figueras_2011,Adam_2012}, which has been widely applied to construct the holographic lattice \cite{Horowitz:2012gs,Horowitz:2012ky,Horowitz:2013jaa,Ling:2013aya,Ling:2013nxa,Donos:2013wia,Ling:2014saa,Andrade:2017leb,Andrade:2017jmt,Andrade:2017ghg,Krikun:2017cyw,Ling:2020qdd,Ling:2023sop}.

In addition, near the boundary $z \to 0$, the asymptotically AdS nature requires that the  fields have the following expansion  behavior:
\begin{equation}\label{eq:7}
    \begin{aligned}
        Q_{tt} & = 1 + q_{tt}(x) z^3 + o\left(z^3\right),      \\
        Q_{zz} & = 1 + o\left(z^3\right),                      \\
        Q_{xx} & = 1 + q_{xx}(x) z^3 + o\left(z^3\right),      \\
        Q_{yy} & = 1 + q_{yy}(x) z^3 + o\left(z^3\right),      \\
        Q_{xz} & = o\left(z\right),                          \\
        A_t    & = \mu_1 - \rho_A(x) z + o\left(z\right),    \\
        B_t    & = \mu_2(x) - \rho_B(x) z + o\left(z\right).
    \end{aligned}
\end{equation}
The asymptotic behavior near $z\rightarrow 0$ also imposes the following conditions:
\begin{equation}\label{eq:qttqxx}
    \begin{aligned}
       q_{tt}(x)+q_{xx}(x)+q_{yy}(x)=0,      \\
       \partial_x q_{xx}(x)=0.
    \end{aligned}
\end{equation}
Equation (\ref{eq:qttqxx}) is equivalent to the fact that the energy-momentum  tensor of the dual field theory is traceless and conserved, which is elaborated in Appendix \ref{appendix:A}.

\section{The CDW Phase with Ionic Lattice}\label{sec3}
\subsection{Perturbation analysis}
In this section, we study the CDW phase in the absence of the superconducting (SC) phase by setting $\eta=0$. First, we briefly review the instability of the standard AdS-RN black brane without lattice, which has previously been investigated in \cite{Ling:2020qdd}. In this case, one  introduces the scalar field $\phi$ and examine the instability under the sinusoidal perturbations with wavenumber $p$ along $ x $-direction, represented by:
\begin{equation}\label{eq:40}
    \delta \Phi(x,z)=\delta \phi^p(z) \cos \left(p x\right).
\end{equation}
Substituting \eqref{eq:40} into the equation of motion for $\Phi$:
\begin{equation}\label{eq:41}
    \nabla^2 \Phi-\frac{1}{4} Z_A^{\prime} F^2-V^{\prime}=0,
\end{equation}
we obtain a linear differential equation for $\delta \phi^p(z)$:
\begin{equation}\label{eq:42}
    \mathcal{O}^{p}(z) \delta \phi^{p}=0.
\end{equation}
Here, $\mathcal{O}^{p}(z)$ is a linear differential operator that depends on $p$. From the original equation, we can see that the elements of the matrix representation of $\mathcal{O}^{p}(z)$ are polynomials in $p$, at most quadratic. This structure allows us to reformulate the problem in two equivalent ways: as an eigenvalue problem, or as a problem of finding the roots of a polynomial in $p$. Specifically, we may discretize Eq. \eqref{eq:42} and then either solve the resulting generalized eigenvalue problem, or directly find roots of $\det [\mathcal{O}^{p}(z)]$. These two approaches are mathematically equivalent and will yield the same values of $p$ for which nontrivial solutions exist.

Next, we intend to examine the instability of the background with ionic lattices. As pointed out in \cite{Andrade:2017leb,Ling:2023sop}, in the presence of the lattice the perturbation form in Eq. \eqref{eq:40} is no longer applicable because the background now exhibits a periodic structure with a wave vector $k$. Instead, one should decompose the wave vector $p$ into periodic regions defined by $\tilde{p} + nk$ with $\tilde{p}\le k$, where $n$ is an integer.

Specifically, we consider the linear perturbation of $\phi$, expressed as:
\begin{equation}\label{eq:43}
    \delta \Phi(x,z)=\delta \phi^{\tilde{p}}(x,z) e^{i \tilde{p} x}.
\end{equation}
Here, $\tilde{p}$ is the wave vector of the CDW, ranging from 0 to $k$. It is noticed that the perturbation mode $\delta \phi^{\tilde{p}}$ with a given wave vector $\tilde{p}$ must also depend on $x$. Consequently, plugging Eq. \eqref{eq:43} into the equation of perturbation \eqref{eq:41} leads to a partial differential equation for $\delta \phi^{\tilde{p}}$, which can be compactly represented as:
\begin{equation}\label{eq:44}
    \mathcal{O}^{\tilde{p}}(x, z) \delta \phi^{\tilde{p}}=0.
\end{equation}
After discretization, the operator $\mathcal{O}^{\tilde{p}}(x, z)$ can again be expressed in matrix form. As the temperature goes down, the system becomes more susceptible to form charge density waves since the thermal fluctuations become weak at lower temperature, allowing for the emergence of ordered states like CDWs. Given a lattice background characterized by a wave vector $k$ and amplitude $\lambda$, its instability can be signaled by the appearance of non-zero eigenvalues for the matrix as one drops down the temperature $T$ to some critical value. For each perturbation mode $\delta \phi^{\tilde{p}}$ with a fixed wave vector $\tilde{p}$, one can identify the critical temperature at which the mode becomes unstable, signaling the onset of CDW formation.

\begin{figure}[htp]
    \centering
    \begin{minipage}{0.5\textwidth}
        \centering
        \includegraphics[width=\textwidth]{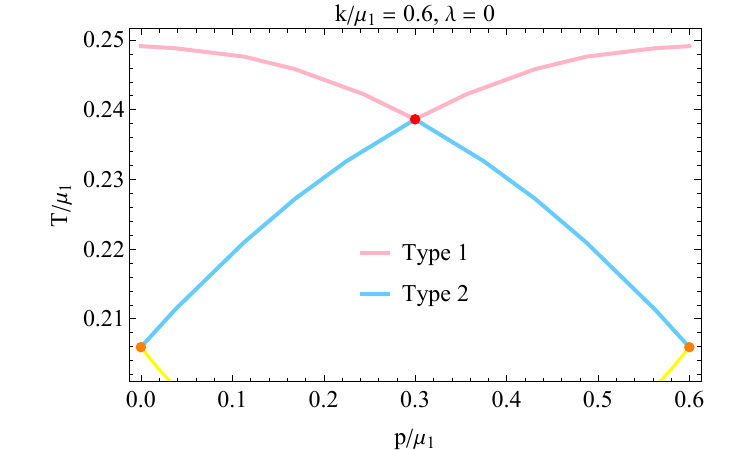}
    \end{minipage}\hfill
    \begin{minipage}{0.5\textwidth}
        \centering
        \includegraphics[width=\textwidth]{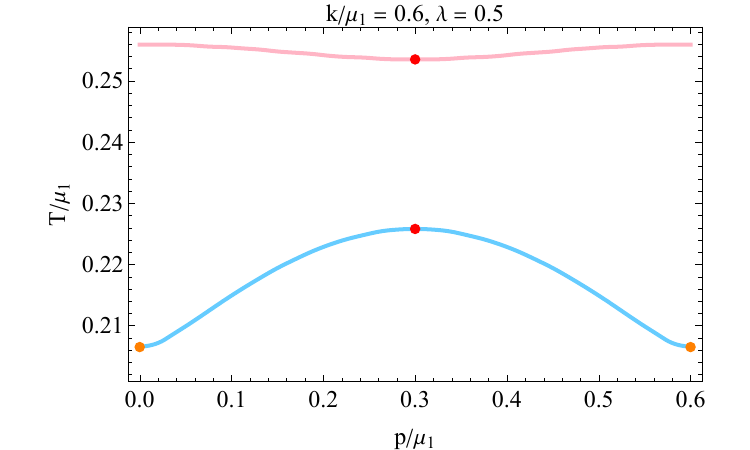}
    \end{minipage}
    \vspace{0.5cm}
    \begin{minipage}{0.5\textwidth}
        \centering
        \includegraphics[width=\textwidth]{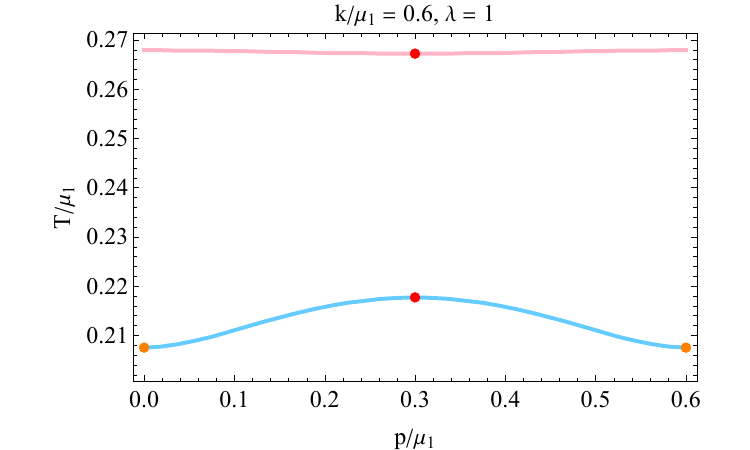}
    \end{minipage}
    \caption{The perturbation phase diagram of CDW with $k/\mu_1 = 0.6$.}
    \label{fig:p-t2}
\end{figure}

\begin{figure}[htp]
    \centering
    \begin{minipage}{0.5\textwidth}
        \centering
        \includegraphics[width=\textwidth]{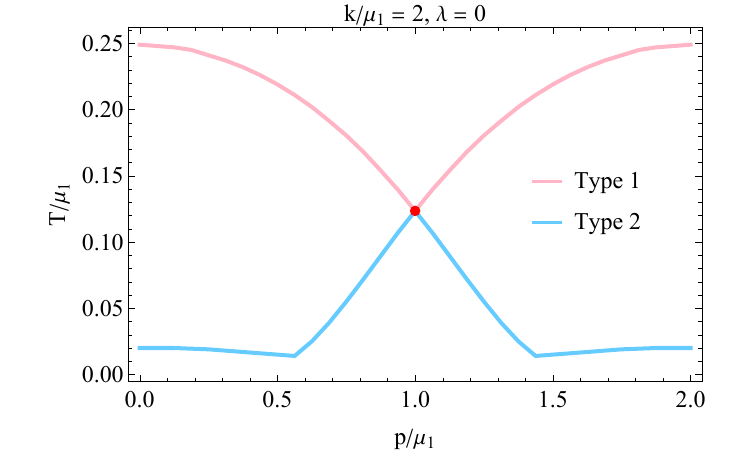} 
    \end{minipage}\hfill
    \begin{minipage}{0.5\textwidth}
        \centering
        \includegraphics[width=\textwidth]{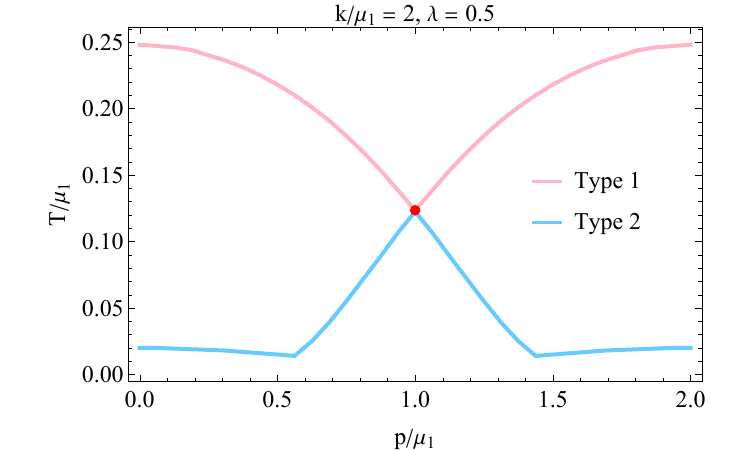} 
    \end{minipage}
       \vspace{0.5cm}
    \begin{minipage}{0.5\textwidth}
        \centering
        \includegraphics[width=\textwidth]{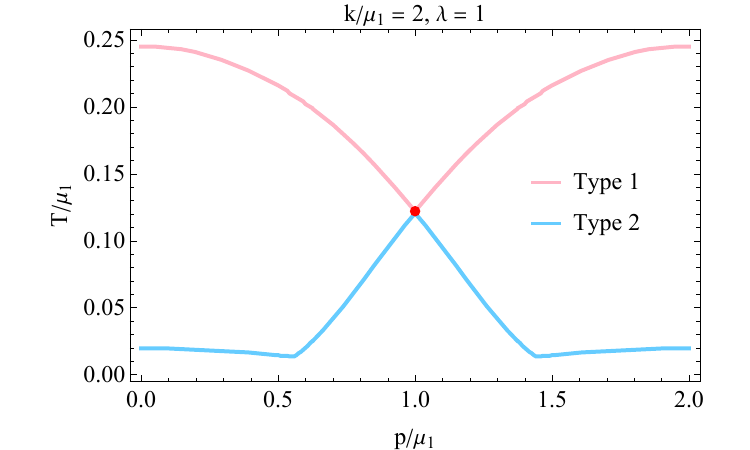}
    \end{minipage}
    \caption{The perturbation phase diagram of CDW with $k/\mu_1 = 2$.
    }
    \label{fig:Instability curve of p-T type 2}
\end{figure}

We perform the numerical analysis based on the above consideration, and then plot the phase diagram in the $\tilde{p}-T$ plane for a given lattice background, as illustrated in Fig. (\ref{fig:p-t2}) and (\ref{fig:Instability curve of p-T type 2}), where the curves denote the critical temperature as the function of $\tilde{p}$. As revealed in \cite{Andrade:2017leb}, the presence of lattice background may change the phase diagram for CDW dramatically, driving it to exhibit a periodic structure in momentum direction as well, which leads to the folding of these curves in the corresponding Brillouin zones. In addition, the number of folding curves decreases with the increase of the lattice wave-vector $k/\mu_1$.  We elaborate this phenomenon based on the model that we are considering, in which distinct types of CDW solutions can be observed below the folding curves. As the lattice wavenumber $k/\mu_1$ is small, specifically exemplified by the case of $k/\mu_1 = 0.6$ in Fig. (\ref{fig:p-t2}), we find that the curves are multiple-folding  but here we are only concerned with the first two orders, which are marked by red dots in the figure. 
The phase diagram implies that in the overlapped region bounded by these two curves there are two distinct solutions for CDW for a given wavenumber $\tilde{p}$. We intend to denote the CDW solution below the curve in pink as Type I solution, while below the curve in blue as Type II solution. With the increase of the lattice amplitude, these two curves become gapped, as illustrated in Fig. (\ref{fig:p-t2}).  In particular, it is worthwhile to point out that in our model, Type I commensurate solution consistently exhibits a lock-in behavior at $\tilde{p}=k$, while Type II solution shows lock-in at $\tilde{p}=k/2$, since at these positions the critical temperature takes the maximal value for the phase transition, independent of the lattice amplitude.  This lock-in behavior is a significant feature different
from those in \cite{Andrade:2017leb}. As the lattice wavenumber $k/\mu_1$ becomes large, specifically exemplified by the case of $k/\mu_1 = 2.0$ in Fig. (\ref{fig:Instability curve of p-T type 2}), the curves are folded only once, marked by a red dot in the figure. It is remarkable to notice that these two curves do not become gapped with the increase of the lattice amplitude (at least in the numerical region that we may approach), which is in contrast to the case with small lattice wavenumber illustrated in Fig. (\ref{fig:p-t2}). Furthermore, we notice that in this case the overlapped region by two curves is negligible such that below the critical temperature one usually finds one type CDW solution, which is locked in at $\tilde{p}=k$.

\begin{figure}[htp]
    \centering
    \includegraphics[width=0.8\linewidth]{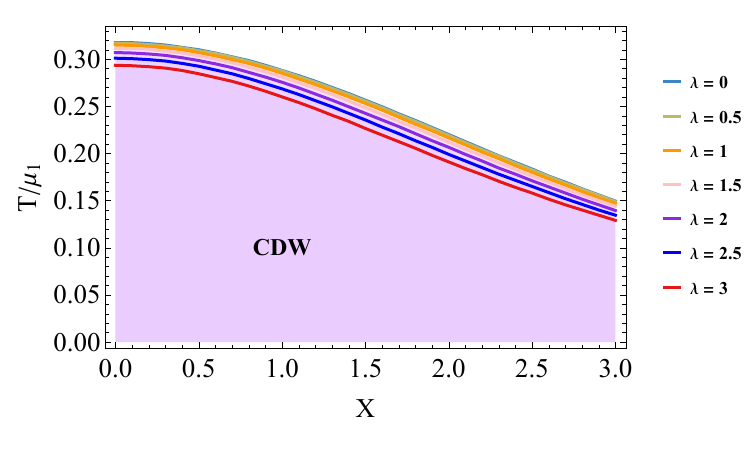}
    \caption{The phase diagram of CDW on $X-T$ plane with various values of the lattice amplitude, where $k/\mu_1=2$ and the commensurate rate $k/\tilde{p}=1$.}
    \label{fig:Phase diagram of CDW}
\end{figure}

Finally, we present the phase diagram of CDW in $X-T$ plane with different amplitudes of ionic lattice in Fig. \ref{fig:Phase diagram of CDW}. It shows that for a fixed doping parameter $ X $, the critical temperature of the phase transition decreases as $\lambda$ increases, implying that the phase transition becomes harder, which is reasonable since deforming the lattice leading to CDW becomes harder as the lattice becomes strong. In addition, one notices that the critical temperature becomes lower with the increase of the doping parameter $X$, which is also understandable since the free carriers increase with the doping, and more free carriers make the transition from a metallic phase to an insulating phase more difficult.

\subsection{Numerical analysis with full back-reaction}

In this subsection, we obtain the background with CDW by solving the equations of motion with full back-reaction, with a focus on the impact of ionic lattice on the condensation of CDW. Firstly, we remark that throughout this paper we will only consider the commensurate states where the period of CDW has a rational relation with the period of lattice background. In this case we expand all the functions with the Fourier modes in $x$ direction and then solve the equations of motion by
pseudo-spectral collocation method and the Newton-Raphson iteration for our numerical calculations \cite{boyd,Dias:2015nua,Andrade:2017jmt}. To guarantee that our solution is not a Ricci soliton, we set the number of collocation points $N_x=30$ and $N_z=50$ throughout the paper, in which case the maximum of $\xi^2$, which measures the discrepancy from the genuine Einstein equations in Einstein-DeTurck method \cite{2010CQGra..27c5002H,Figueras_2011,Adam_2012,Horowitz:2012gs,Horowitz:2012ky,Horowitz:2013jaa,Donos:2013wia}, is less than $5\times 10^{-9}$ and the maximum reduces with the increase of the collocation points. Since the phase diagram in the previous subsection reveals that there are distinct types of solutions below the critical temperature, we begin by justifying this observation by solving the equations of motion with full back-reactions. 

\subsubsection{Two types of CDW solutions}
\begin{figure}[htp]
    \centering
    \begin{minipage}{0.5\textwidth}
        \centering
        \includegraphics[width=\textwidth]{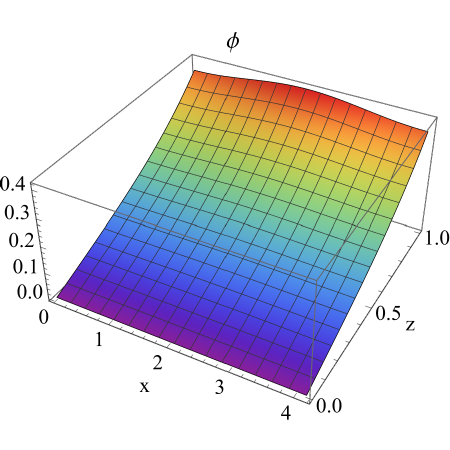} 
    \end{minipage}\hfill
    \begin{minipage}{0.5\textwidth}
        \centering
        \includegraphics[width=\textwidth]{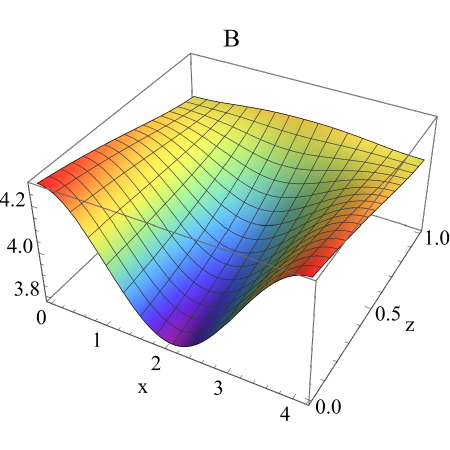} 
    \end{minipage}
    \caption{Type I: Three-dimensional numerical simulations of scalar field $\phi$ (left) and gauge field $B$ (right) for parameters $k/\mu_1 = 0.6$, $T/\mu_1 = 0.2$, and $\lambda = 0.1$. The $x$ and $y$ axes represent spatial coordinates, while the $z$-axis shows the field magnitude.
    }
    \label{fig:CDWtype1}
\end{figure}

\begin{figure}[htp]
    \centering
    \begin{minipage}{0.5\textwidth}
        \centering
        \includegraphics[width=\textwidth]{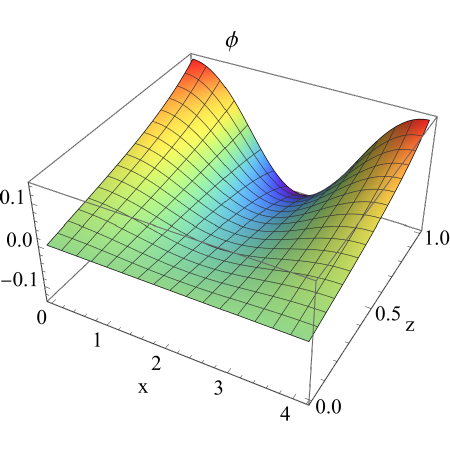} 
    \end{minipage}\hfill
    \begin{minipage}{0.5\textwidth}
        \centering
        \includegraphics[width=\textwidth]{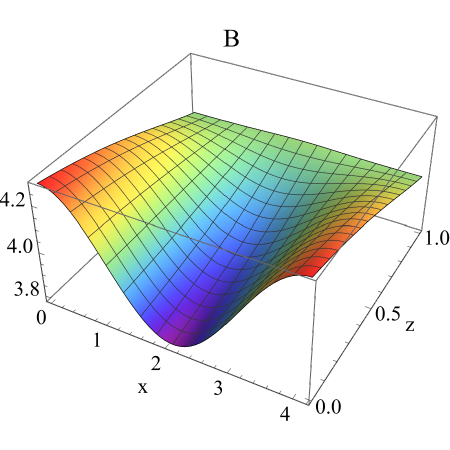} 
    \end{minipage}
    \caption{Type II: Three-dimensional numerical simulations of scalar field $\phi$ (left) and gauge field $B$ (right) for parameters $k/\mu_1 = 0.6$, $T/\mu_1 = 0.2$, and $\lambda = 0.1$. The $x$ and $y$ axes represent spatial coordinates, while the $z$-axis shows the field magnitude.
    }
    \label{fig:CDWtype2}
\end{figure}

As indicated by the phase diagram, we numerically solve the equations of motion to demonstrate two distinct background solutions at the same temperature below the critical ones. These solutions are illustrated in Fig. (\ref{fig:CDWtype1}) and Fig. (\ref{fig:CDWtype2}).
Both Type I and Type II solutions exhibit a charge density wave with a periodicity matching that of the underlying lattice. In addition,  it is noticed that the magnitude of the scalar  field $\phi$ in Type I solution is much greater than that of the scalar  field $\phi$ in Type II solution. This is understandable since the temperature is closer to the critical temperature of Type II. In order to compare their relative stability, we compute their average free energy which is given by the following thermal quantities
\begin{equation}
    F=m-\mu_1 Q_A-\mu_2 Q_B-T S,
\end{equation}
where
\begin{equation}
    \begin{aligned}
        m   & =4+\frac{\mu_1^2}{4}+\frac{\mu_2^2}{4}+\frac{6 k_c}{2 \pi} \int_0^{2 \pi / k_c} q_{xx}(x) d x+\frac{6 k_c}{2 \pi} \int_0^{2 \pi / k_c} q_{yy}(x) d x, \\
        Q_A & =  \frac{ \mu_1 k_c}{2 \pi} \int_0^{2 \pi / k_c} \rho_A(x) d x,                          \\
        Q_B & =  \frac{  \mu_2 k_c}{2 \pi} \int_0^{2 \pi / k_c} \rho_B(x) d x,                         \\
        S   & =k_c \int_0^{2 \pi / k_c} \sqrt{Q_{x x}(x, 1) Q_{y y}(x, 1)} d x .
    \end{aligned}
\end{equation}
$m$ is the averaged mass density of the black hole and we explain it in Appendix \ref{appendix:A}. Fig. (\ref{fig:f-t2}) illustrates the free energy of two types of CDW solutions as well as lattice background without CDW, revealing several key observations. First, both CDW solutions exhibit lower free energy than the lattice background, indicating a phase transition below the critical temperature. Second, Type I CDW solution demonstrates a higher critical temperature and lower free energy than Type II at the same temperature, making it the most stable configuration. In contrast, Type II appears to be an intermediate, allowable but unstable phase. In addition, we notice that for Type II solutions, the free energy of the commensurate state with $\tilde{p}/k=1/2$ is lower than that of state with $\tilde{p}/k=1/1$, thus expecting that Type II solutions are locked-in at $\tilde{p}/k=1/2$, which is consistent with the phase diagram in  Fig. (\ref{fig:p-t2}). Given these findings, subsequent sections will concentrate on the transport behavior of Type I CDW solutions, as they represent the most stable and relevant configuration for further analysis.

\begin{figure}[htp]
    \centering
    \includegraphics[width=0.8\linewidth]{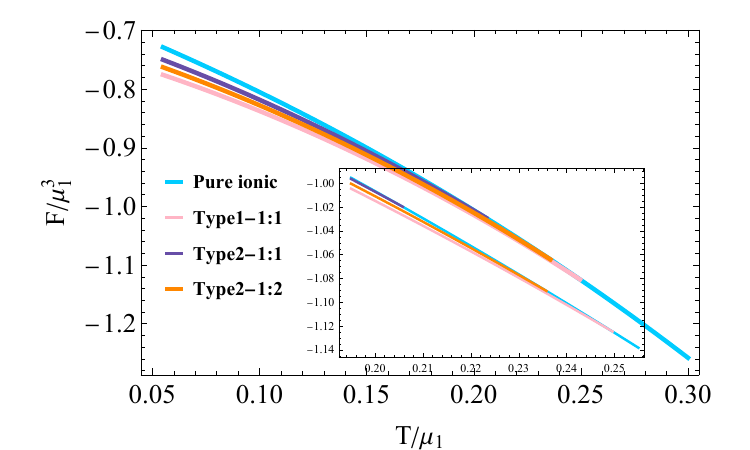}
    \caption{ The average free energy for two types CDW solutions and lattice background, where the commensurate ratio is $\tilde{p}:k$ and we set $k/\mu_1=0.6$ here.}
    \label{fig:f-t2}
\end{figure}

For larger values of $k/\mu_1$, such as $k/\mu_1=2$ in Fig. (\ref{fig:Instability curve of p-T type 2}), only Type I solutions are observed below the critical temperature.

Due to its higher critical temperature and lower free energy, we focus on Type I solution with $ k/\mu_1 = 2 $ in the subsequent analysis. 

\subsubsection{The dilaton field $\phi$}
\begin{figure}[htp]
    \centering
    \begin{minipage}{0.5\textwidth}
        \centering
        \includegraphics[width=\textwidth]{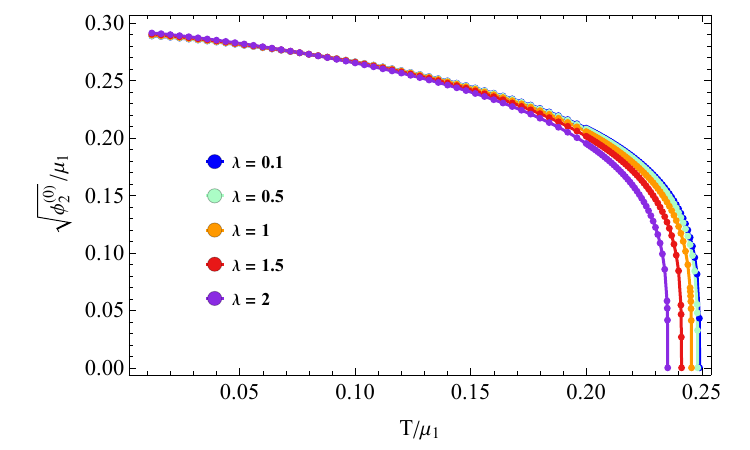}
    \end{minipage}\hfill
    \begin{minipage}{0.5\textwidth}
        \centering
        \includegraphics[width=\textwidth]{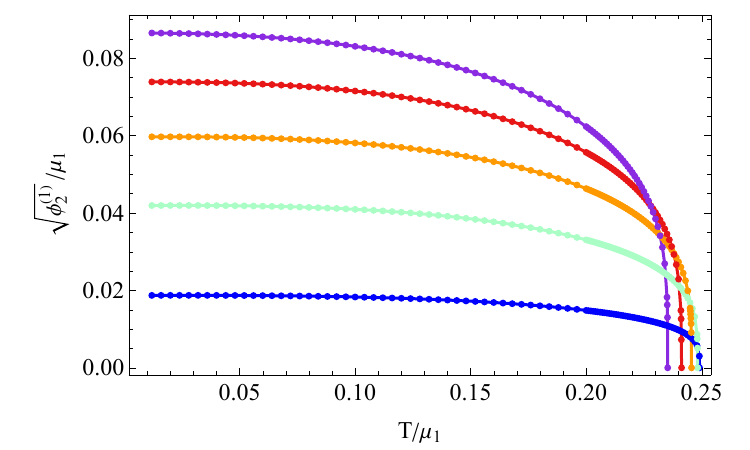}
    \end{minipage}
    \caption{The leading orders of $\Phi$ as the function of temperature T with various values of the lattice amplitude, where the doping parameter is fixed at $X=1.6$.}
    \label{fig:The zeroth to the third term of Phi as the function of temperature T.}
\end{figure}

\begin{figure}[htp]
    \centering
    \begin{minipage}{0.5\textwidth}
        \centering
        \includegraphics[width=\textwidth]{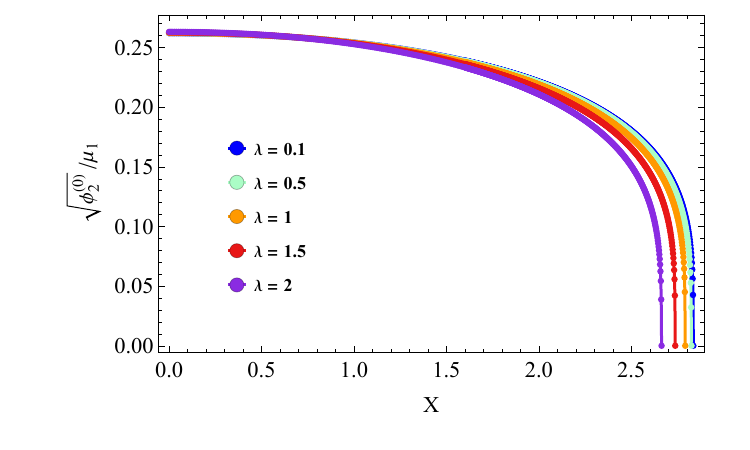}
    \end{minipage}\hfill
    \begin{minipage}{0.5\textwidth}
        \centering
        \includegraphics[width=\textwidth]{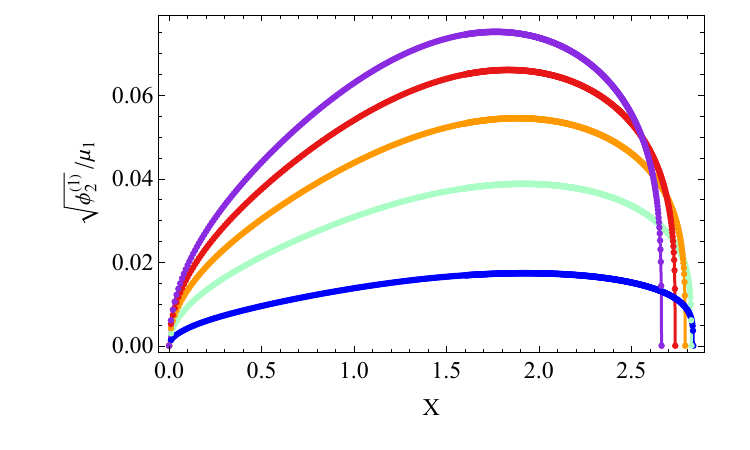}
    \end{minipage}
    \caption{The leading orders of $\Phi$ as the function of doping parameter $X$ with various values of the lattice amplitude, where the temperature is fixed at $T/\mu_1=0.16$.}
    \label{fig:The zeroth to the third term of Phi as the function of doping parameter x.}
\end{figure}

Since the generation of CDW in the background is signaled by the non-zero solution of the dilaton field $\phi$, next we intend to look into its behavior during the course of phase transition.
In the asymptotically AdS spacetime, the equation of motion for   $\phi(z)$ requires that near the boundary $z=0$ it behaves as,
\begin{equation}
    \phi \approx  \phi_1 z+ \phi_2 z^2+\cdots.
\end{equation}
To guarantee the translational symmetry is broken in a spontaneous manner, we turn off the source term in numerical analysis by setting $\phi_1=0$. Upon numerically solving the equations of motion with full back-reaction, we find that in the presence of ionic lattice with wavenumber $k$, the commensurate solution of the dilaton field $\phi_2$ with $\tilde{p}/k=1$ exhibits the following expansion behavior along $x$ direction:
$$
    \phi_2(x)=\phi_2^{(0)}+\phi_2^{(1)} \cos \left(k x\right)+\phi_2^{(2)} \cos \left(2 k x\right)+\phi_2^{(3)} \cos \left(3 k x\right) +\cdots.
$$
Notably, this result is in contrast to the expansion previously obtained in \cite{Ling:2020qdd}, where the lattice background is absent and the expansion contains the odd terms only:
$$
    \phi_2(x)=\phi_2^{(1)} \cos \left(k_c x\right)+\phi_2^{(3)} \cos \left(3 k_c x\right) +\cdots.
$$
This discrepancy highlights that, in the presence of an explicit ionic lattice, the Fourier modes of $\phi$ solution can be any integer multiple of the wave vector of the ionic lattice, resulting in a period identical to that of the lattice. The concrete numerical results for the first two leading orders of the coefficients are shown in Fig. (\ref{fig:The zeroth to the third term of Phi as the function of temperature T.}) and Fig. (\ref{fig:The zeroth to the third term of Phi as the function of doping parameter x.}).

Fig. (\ref{fig:The zeroth to the third term of Phi as the function of temperature T.}) illustrates the condensate behavior of the order parameter as a function of temperature. It is obvious to see that around the critical temperature the system undergoes a second-order phase transition, namely from a metallic phase to an insulating CDW phase. With the increase of the lattice amplitude, the critical temperature becomes lower, which is consistent with the phase diagram previously obtained by perturbative analysis. In addition, as the temperature goes down, all the coefficients become saturated. It is interesting to notice that the value of the saturation for the leading order $\phi_2^{(0)}$ does not change much with the variation of the lattice amplitude, while for the subleading order $\phi_2^{(1)}$  the saturation value increases evidently with the lattice amplitude $\lambda$. 

Fig. (\ref{fig:The zeroth to the third term of Phi as the function of doping parameter x.}) demonstrates the order parameters as a function of the doping parameter $X$ while the temperature is fixed below the critical $T_c$. It is interesting to note that except for the zeroth coefficient $\phi_2^{(0)}$, which exhibits saturation behavior as $X\rightarrow0$, all the odd terms tend to 0 as $X\rightarrow0$. Thus, these coefficients take the maximal value at some $X$.

\subsubsection{The charge density $\rho_B$}
We now examine the behavior of the charge density $\rho_B(x)$ as a function of the temperature $T$, doping parameter $X$, as well as the lattice amplitude $\lambda$. The charge density in the dual theory can be extracted from the bulk solution of $B_t(x,z) = \mu(x) - \rho_B (x)z + \cdots$.

In the presence of an ionic lattice with $\tilde{p}/k=1$, our numerical results reveal that all cosine modes contribute to the Fourier expansion of the charge density:
\begin{equation}
    \rho_B(x) = \rho_B^{(0)} + \rho_B^{(1)}\cos(kx) + \rho_B^{(2)}\cos(2kx) + \rho_B^{(3)}\cos(3kx) + \cdots.
\end{equation}
Conversely, in the absence of an ionic lattice, it has been demonstrated in \cite{Ling:2020qdd} that only even-order terms appear in the expansion:
\begin{equation}
    \rho_B(x) = \rho_B^{(0)} + \rho_B^{(2)}\cos(2k_cx) + \cdots.
\end{equation}

More importantly, we stress that $\rho_B(x)$ contains the total charge density in the dual system. Due to the presence of ionic lattice, it already exhibits a periodic behavior along $ x $-direction before the critical temperature of CDW, which is in contrast to the case without a lattice background. Therefore, to isolate the contribution of the CDW after the phase transition, we define the difference of the charge density $\Delta \rho_B(x)$ by subtracting the charge density of the lattice background without CDW at the same temperature, which has previously been performed in \cite{Andrade:2017ghg,Ling:2023sop},

\begin{equation}
    \Delta \rho_B^{\text{CDW}}(x) = \left| \rho_B(x) - \rho_B^{\text{Lattice}}(x) \right|.
\end{equation}

\begin{figure}[htp]
    \centering
    \begin{minipage}{0.7\textwidth}
        \centering
        \includegraphics[width=\textwidth]{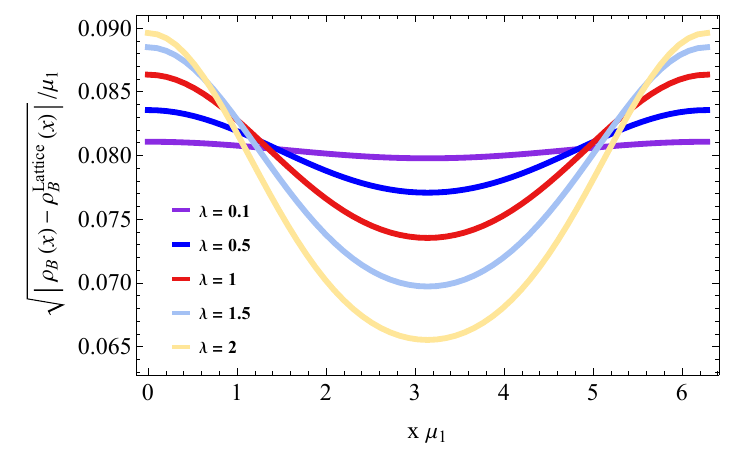}
    \end{minipage}
    \caption{The magnitude of $\sqrt{\Delta \rho_B^{\text{CDW}}(x)}$ of $T/\mu_1=0.168$ with different $\lambda$, where the doping parameter $X=1.6$.  We plot the diagram in the same period of the ionic lattice.}
    \label{fig:deltarhocdw}
\end{figure}

Fig. (\ref{fig:deltarhocdw}) plots the spatial distribution of $\Delta \rho_B^{\text{CDW}}(x)$ as the charge density of CDW along $x$ direction for $T/\mu_1 = 0.168$ ($T_c/\mu_1 =$ 0.2491, 0.2483, 0.2456, 0.2413 and 0.2353 correspondingly) with various $\lambda$ and $X$. The results indicate that the magnitude of $\Delta \rho_B^{\text{CDW}}(x)$ becomes larger with the increase of the lattice amplitude, while its average value becomes smaller with the increase of the doping parameter $X$.

Fourier analysis of $\Delta \rho_B^{\text{CDW}}(x)$ reveals the presence of all orders:
\begin{equation}
    \Delta \rho_B^{\text{CDW}}(x) = \Delta\rho_B^{(0)} + \Delta\rho_B^{(1)} \cos(kx) + \Delta\rho_B^{(2)} \cos(2kx) + \Delta\rho_B^{(3)} \cos(3kx) + \cdots. 
\end{equation}

The presence of all the orders in the Fourier modes of the charge density could be treated as an improvement in a series of work on the construction of holographic CDW within two-gauge formalism \cite{Ling:2014saa,Ling:2019gjy,Ling:2020qdd}. In \cite{Ling:2014saa,Ling:2019gjy} only odd orders of CDW appear while in \cite{Ling:2020qdd} only even orders of CDW appear, thus distinct behavior of the CDW and its relation with SC are observed.  Specifically, in \cite{Ling:2019gjy}  we show that in the absence of  free charges $\rho_B^{(0)}$, the superconductivity can be induced by the presence of CDW $\rho_B^{(1)}$. While in  \cite{Ling:2020qdd} we demonstrate  that the critical temperature of SC phase is not sensitive to the presence of even orders of CDW.  In current paper we demonstrate that all the orders of CDW may appear due to the presence of the ionic lattice. 

\begin{figure}[htp]
    \centering
    \begin{minipage}{0.5\textwidth}
        \centering
        \includegraphics[width=\textwidth]{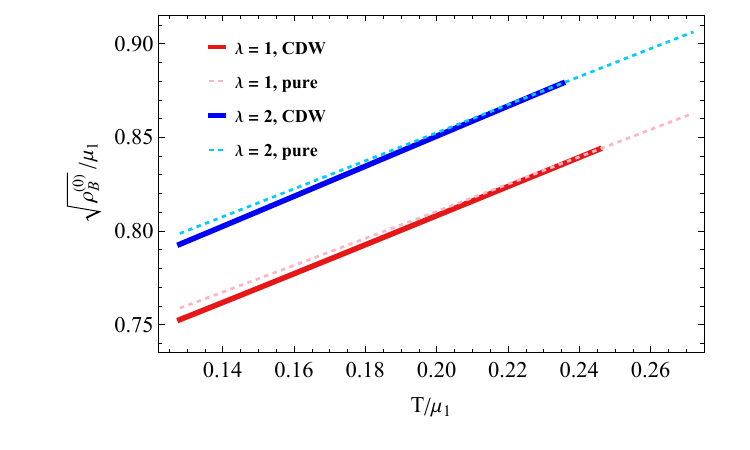} 
    \end{minipage}\hfill
    \begin{minipage}{0.5\textwidth}
        \centering
        \includegraphics[width=\textwidth]{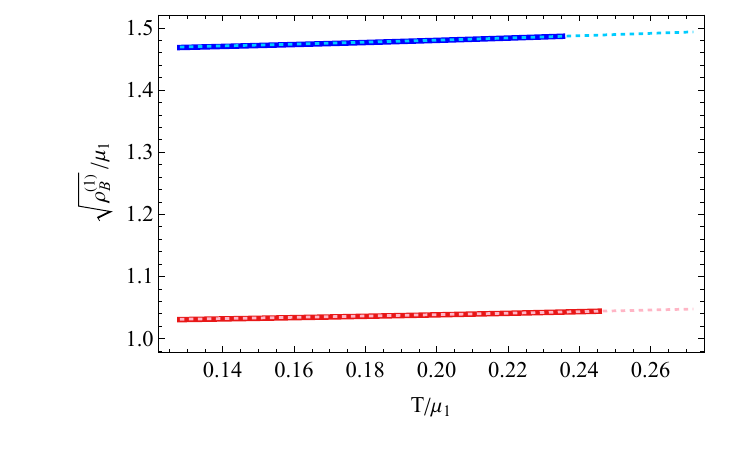} 
    \end{minipage}
    \vspace{0.5cm} 
    \begin{minipage}{0.5\textwidth}
        \centering
        \includegraphics[width=\textwidth]{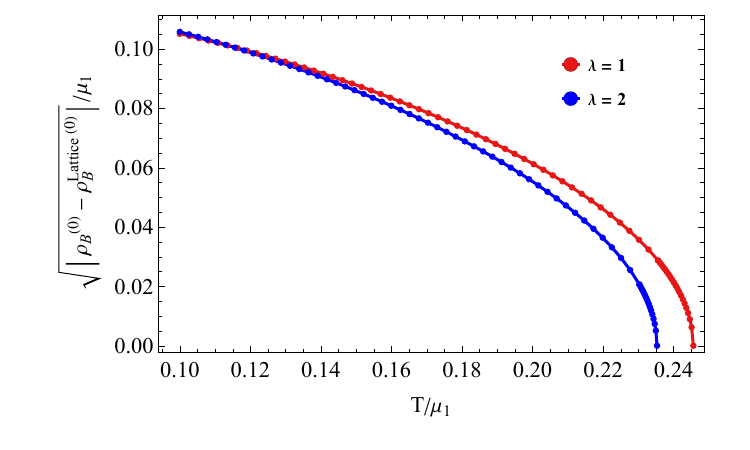} 
    \end{minipage}\hfill
    \begin{minipage}{0.5\textwidth}
        \centering
        \includegraphics[width=\textwidth]{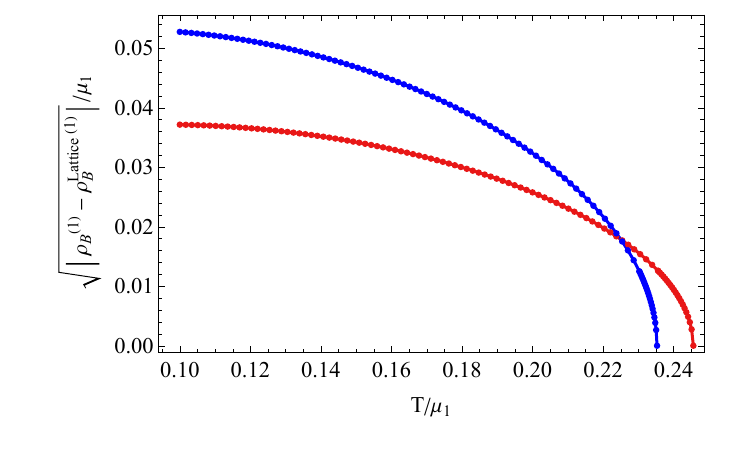} 
    \end{minipage}
    \caption{The leading orders of $\rho_B(x)$ and $\Delta \rho_B^{\text{CDW}}(x)$  as the function of temperature $T$, where the doping parameter $X=1.6$.}
    \label{fig:phi The zeroth to the third term of rho T}
\end{figure}

\begin{figure}[htp]
    \centering
    \begin{minipage}{0.5\textwidth}
        \centering
        \includegraphics[width=\textwidth]{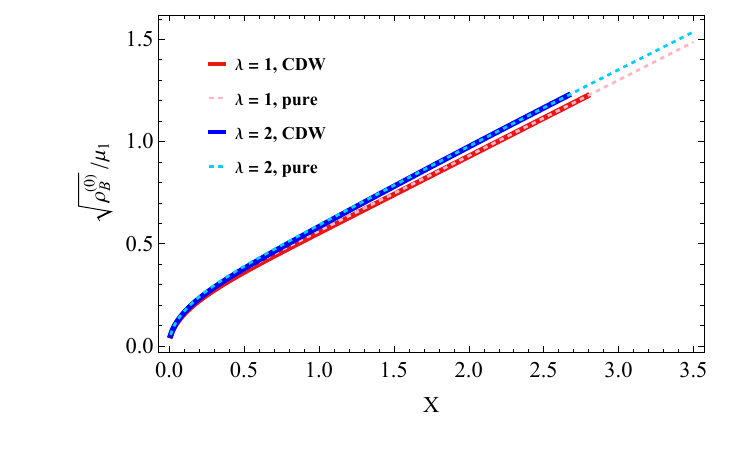} 
    \end{minipage}\hfill
    \begin{minipage}{0.5\textwidth}
        \centering
        \includegraphics[width=\textwidth]{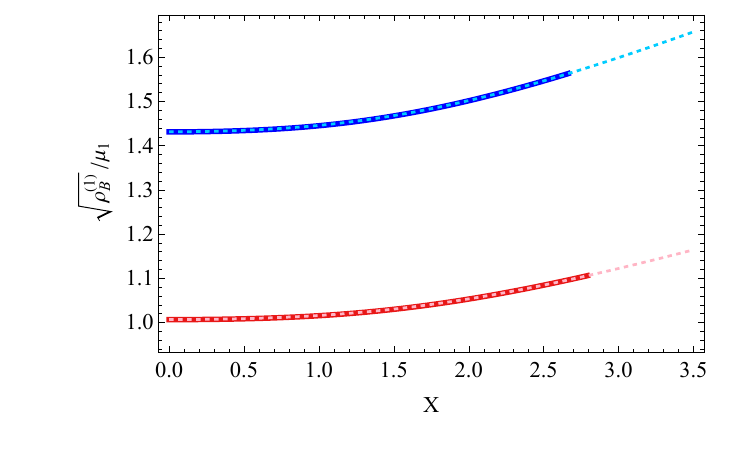} 
    \end{minipage}
    \vspace{0.5cm} 
    \begin{minipage}{0.5\textwidth}
        \centering
        \includegraphics[width=\textwidth]{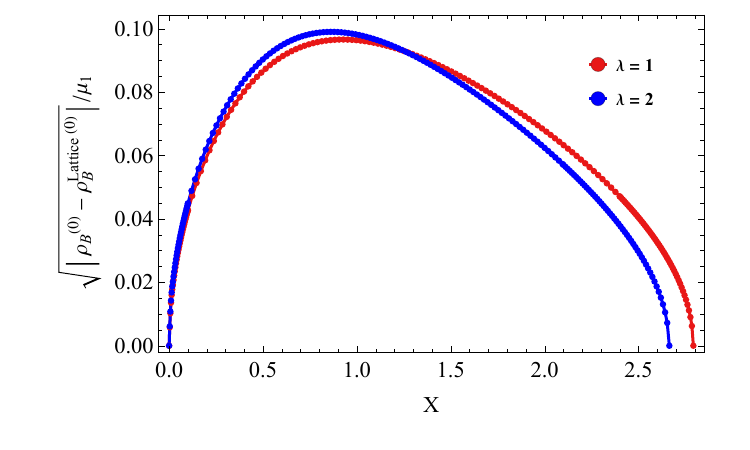} 
    \end{minipage}\hfill
    \begin{minipage}{0.5\textwidth}
        \centering
        \includegraphics[width=\textwidth]{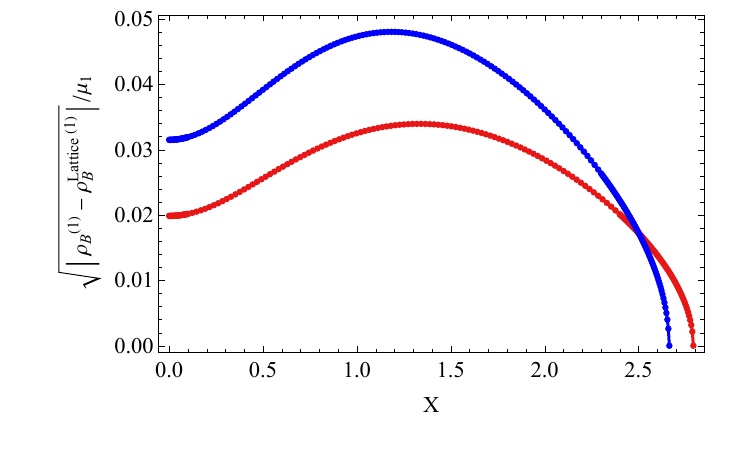} 
    \end{minipage}
    \caption{The leading orders of $\rho_B(x)$ and $\Delta \rho_B^{\text{CDW}}(x)$  as the function of doping parameter X, where the temperature  $T/\mu_1=0.16$.}
    \label{fig:phi The zeroth to the third term of rho X}
\end{figure}

Next we investigate  the dependence of the charge density on temperature and doping parameter through numerical simulations, and the results are illustrated in Fig. (\ref{fig:phi The zeroth to the third term of rho T}) and Fig. (\ref{fig:phi The zeroth to the third term of rho X}), respectively.

Our analysis of the charge density behavior reveals several key findings. Fig. (\ref{fig:phi The zeroth to the third term of rho T}) demonstrates that for a fixed doping parameter $X$ and lattice amplitude $\lambda$, the leading orders of the charge density $\rho_B(x)$ decrease as the temperature goes down. On the contrary, $\Delta \rho_B^{\text{CDW}}(x)$ which evaluates the contribution from CDW becomes larger. This tendency is reasonable since the magnitude of CDW is expected to become larger as the temperature is further dropping down below the critical temperature $T_c$. Similarly, for fixed $X$ and $T/\mu_1$, the increase of the lattice amplitude $\lambda$ leads to an increase in all orders of $\rho_B(x)$, but a decrease in the leading order of $\Delta \rho_B^{\text{CDW}}(x)$, implying that the formation of CDW becomes harder as the lattice becomes stronger. All the above observations indicate that lower temperature and weak lattice may enhance the formation of CDW.

Fig. (\ref{fig:phi The zeroth to the third term of rho X}) plots $\rho_B(x)$ as well as $\Delta \rho_B^{\text{CDW}}(x)$ as a function of the doping parameter $X$, where the temperature is fix at $T/\mu_1=0.16$. Without surprise the total charge density $\rho_B(x)$ monotonously increases with the doping parameter $X$, but it is interesting to notice that for $\Delta \rho_B^{\text{CDW}}(x)$, there exists an optimal doping where it takes the maximal value, indicating the highest contribution of CDW.

\section{The Superconducting Phase with ionic lattice}\label{sec4}
In this section, we investigate the SC phase without the CDW phase by setting $\phi = 0$ throughout. In parallel, we will obtain the phase diagram in $T-X$ plane by evaluating the critical temperature under the perturbative analysis, and then obtain the background with superconducting condensate by solving the equations of motion with backreactions.

\subsection{Perturbation analysis}

\begin{figure}[htp]
    \centering
    \includegraphics[width=0.7\linewidth]{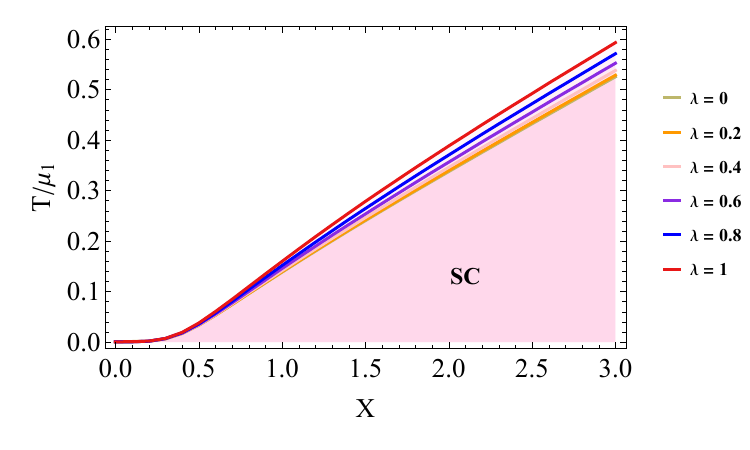}
    \caption{The phase diagram of SC in the $X-T$ plane for various values of the lattice amplitude under the commensurate state with $\tilde{p}/k=1$.}
    \label{fig:SC phase diagram}
\end{figure}

Over a background with ionic lattice, the equation of motion for $\eta$ is,
\begin{equation}\label{etaequation}
    \nabla^2 \eta-m_v^2 \eta-(e B)^2 \eta=0.
\end{equation}
Similar to the treatment of CDW, we consider the perturbation modes as follows,
\begin{equation}\label{etaperturb}
    \delta\eta(x,z)= \delta\eta^{\tilde{p}}(z) e^{i \tilde{p} x} .
\end{equation}
By substituting \eqref{etaperturb} into Eq. \eqref{etaequation}, the linear perturbation equation can be represented as,
\begin{equation}\label{eq:44v2}
    \mathcal{O}^{\tilde{p}}(x, z) \delta \eta^{\tilde{p}}=0.
\end{equation}
Similarly, the issue on the existence of nonzero solution for $\delta \eta^{\tilde{p}}$ can be  changed into an issue on the eigenvalue problem of the operator $\mathcal{O}^{\tilde{p}}(x, z)$. For simplicity, throughout this paper we will only consider the commensurate state with $\tilde{p}/k=1$ for SC phase. Setting $m_v^2 = -8$ and $e = 4$, we obtain the critical temperature as the function of the doping parameter $X$ and plot the phase diagram on $X-T$ plane for various lattice amplitude in Fig. (\ref{fig:SC phase diagram}).

Fig. (\ref{fig:SC phase diagram}) reveals two features of the SC critical behavior. 
First, the critical temperature for SC phase transition goes up with the increase of doping parameter $X$, which is favorable as higher doping levels produce more carriers for Cooper pair formation, facilitating condensation. This aligns with previous findings for lattice-free backgrounds \cite{Ling:2020qdd}. Second, for a fixed $X$, the critical temperature becomes higher  with the increase of the lattice amplitude $\lambda$, which is contrary to the behavior of CDW. This inverse relationship occurs because larger ionic potentials inhibit spontaneous breaking of translational symmetry while promoting superconducting condensation. 

These observations provide valuable insights into how the interplay between doping and lattice potential affects the critical temperature in superconducting systems.

\subsection{Numerical analysis with full back-reaction}
In this subsection we obtain the background with superconducting condensation by solving the equations of motion with full back-reaction, and then discuss the behavior of the order parameter $\eta$ as well as  the charge density $\rho_B(x)$ with various values of the doping parameter $X$ and the lattice amplitude $\lambda$. One numerical solution for the field $\eta$ and the gauge field  $B(x)$ is shown as below in Fig. (\ref{fig:SCnumerical_result}). We remark that since $\eta$ is the magnitude of the complex field, $\eta > 0$ is guaranteed for the solution as illustrated in this figure. Thus it gives rise to genuine $U(1)$ gauge symmetry breaking rather than the Stuckelberg mechanism.
\begin{figure}[htp]
    \centering
    \begin{minipage}{0.5\textwidth}
        \centering
        \includegraphics[width=\textwidth]{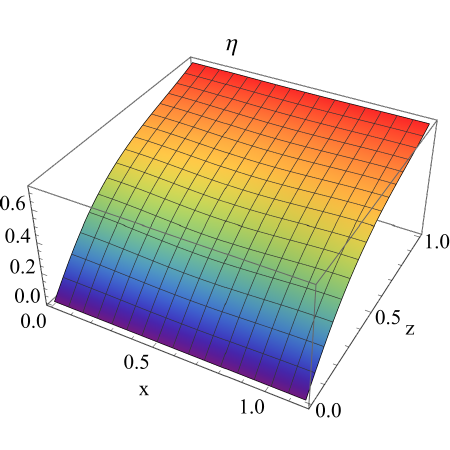} 
    \end{minipage}\hfill
    \begin{minipage}{0.5\textwidth}
        \centering
        \includegraphics[width=\textwidth]{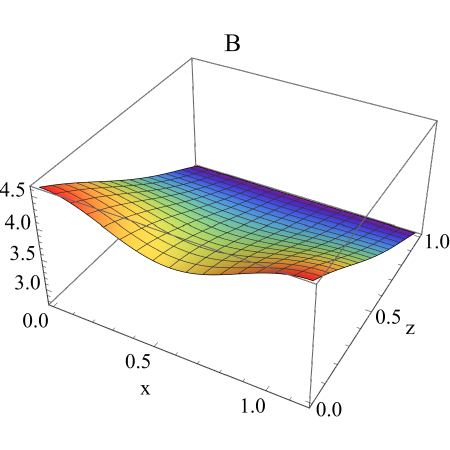} 
    \end{minipage}
    \caption{Numerical results of the field $\eta$ and $B$ with the parameters $k/\mu_1$=0.6, $T/\mu_1$=0.16 and $\lambda=0.1$.}
    
    \label{fig:SCnumerical_result}
\end{figure}

\subsubsection{The behavior of the condensation $\eta$}
Similarly, the requirement of the asymptotic AdS spacetime gives rise to the expansion of $\eta(z)$ near the boundary ($z\to 0$) as,
\begin{equation}
    \eta \approx  \eta_1 z+ \eta_2 z^2+\cdots.
\end{equation}
We need to set $\eta_1=0$ at $z=0$ to guarantee that U(1) symmetry is broken spontaneously. After numerically solving all the equations of motion with full back-reaction, we find $\eta_2$ behaves as
\begin{equation}
    \eta_2(x)=\eta_2^{(0)}+\eta_2^{(1)} \cos \left(k x\right)+\eta_2^{(2)} \cos \left(2 k x\right)+\eta_2^{(3)} \cos \left(3 k x\right) \cdots,
\end{equation}
\begin{figure}[htp]
    \centering
    \begin{minipage}{0.5\textwidth}
        \centering
        \includegraphics[width=\textwidth]{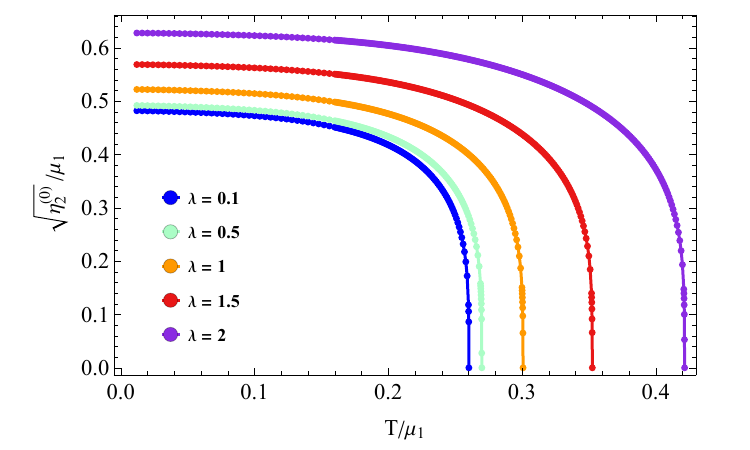} 
    \end{minipage}\hfill
    \begin{minipage}{0.5\textwidth}
        \centering
        \includegraphics[width=\textwidth]{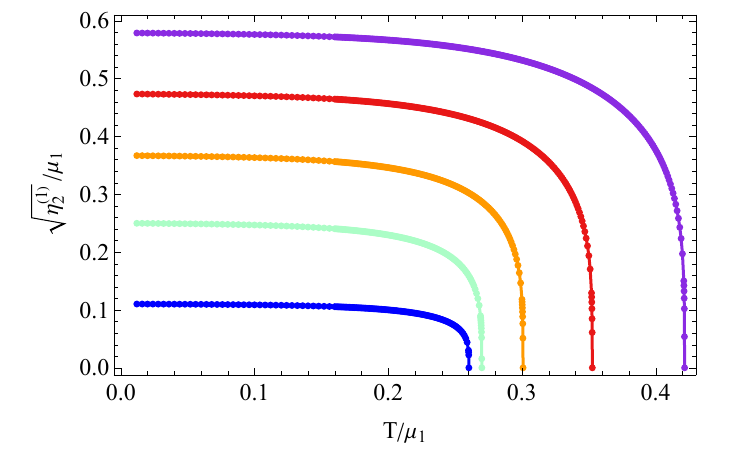} 
    \end{minipage}
    \caption{The leading orders of $\eta$ as the function of temperature $ T $ for various values of the lattice amplitude, where the doping parameter is fixed as $X=1.6$.}
    \label{fig:The zeroth to the eta third term of eta as the function of temperature T.}
\end{figure}

\begin{figure}[htp]
    \centering
    \begin{minipage}{0.5\textwidth}
        \centering
        \includegraphics[width=\textwidth]{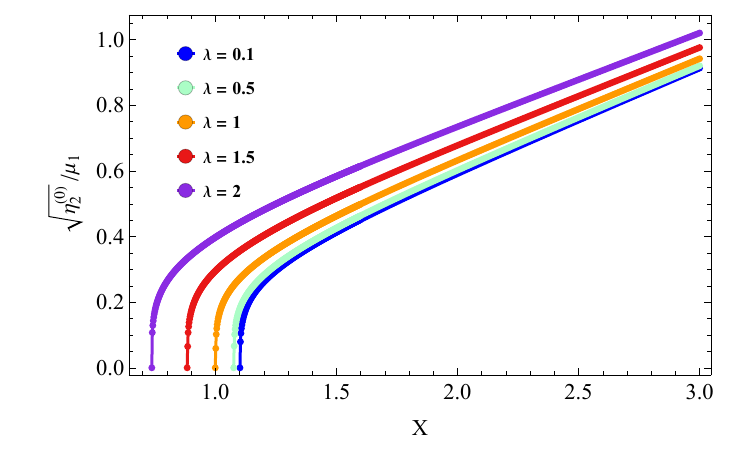} 
    \end{minipage}\hfill
    \begin{minipage}{0.5\textwidth}
        \centering
        \includegraphics[width=\textwidth]{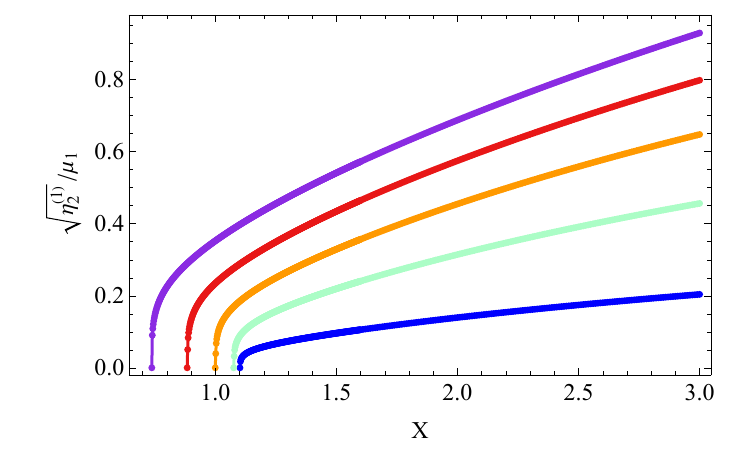} 
    \end{minipage}
    
    \caption{The zeroth and the first order of $\eta$ as the function of doping parameter $ X $ with various values of the lattice amplitude, where the temperature is fixed as $T/\mu_1=0.16$.}
    \label{fig:The zeroth to the third term of eta as the function of temperature X.}
\end{figure}

Similar to the behavior of $\phi$ field discussed before, the Fourier modes of $\eta$ contain integer multiple of the wave vector of the ionic lattice, so characterized by the same period of the lattice background. The concrete numerical results are shown in Fig. (\ref{fig:The zeroth to the eta third term of eta as the function of temperature T.}) and (\ref{fig:The zeroth to the third term of eta as the function of temperature X.}).

Fig. (\ref{fig:The zeroth to the eta third term of eta as the function of temperature T.}) illustrates the temperature behavior of the leading orders of $\eta$ in the Fourier expansion after the occurrence of  the phase transition. It is seen that the emergence of superconductivity is a typical second order phase transition. We denote $\eta_2^{(0)}$ as the order parameter of superconductivity. For a fixed doping parameter, it is observed that the critical temperature  goes up with the increase of $\lambda$. In addition, the saturated value of the order parameter at $T\rightarrow 0$ becomes higher as well. These results explicitly indicate that superconducting condensation benefits from the ionic lattice background in holographic approach, which is also consistent with the phenomenon observed in laboratory. 

Fig. (\ref{fig:The zeroth to the third term of eta as the function of temperature X.}) demonstrates the doping dependence of the condensate field. In general, the magnitude of these coefficients becomes larger with the increase of the doping parameter. In addition, as the lattice amplitude becomes larger, the increase of the SC condensation 
becomes more obvious.

\subsubsection{The behavior of the charge density $\rho_B$}
Now we turn to study the behavior of charge density $\rho_B(x)$ after the SC phase transition with different $X$, $T$ and $\lambda$. Similarly, in the presence of ionic lattice we find all the series appear in the Fourier expansion of the numerical solution. 
\begin{equation}
    \rho_B(x)=\rho_B^{(0)}+\rho_B^{(1)} \cos \left(k x\right)+\rho_B^{(2)} \cos \left(2 k x\right)+\rho_B^{(3)} \cos \left(3 k x\right) \cdots.
\end{equation}
In particular, to describe the discrepancy of the charge density before and after phase transition, we may define $\Delta \rho_B(x)$ as:
\begin{equation}
    \Delta \rho_B^{\text{SC}}(x) = \left| \rho_B(x) - \rho_B^{\text{Lattice}}(x) \right|,
\end{equation}
which are numerically obtained at the same temperature. $\Delta \rho_B^{\text{SC}}(x)$ can also be expressed as a Fourier series:
\begin{equation}
    \Delta \rho_B^{\text{SC}}(x)=\Delta\rho_B^{(0)}+\Delta\rho_B^{(1)} \cos \left(k x\right)+\Delta\rho_B^{(2)} \cos \left(2 k x\right)+\Delta\rho_B^{(3)} \cos \left(3 k x\right) + \cdots.
\end{equation}

\begin{figure}[htp]
    \centering
    \begin{minipage}{0.49\textwidth}
        \centering
        \includegraphics[width=\textwidth]{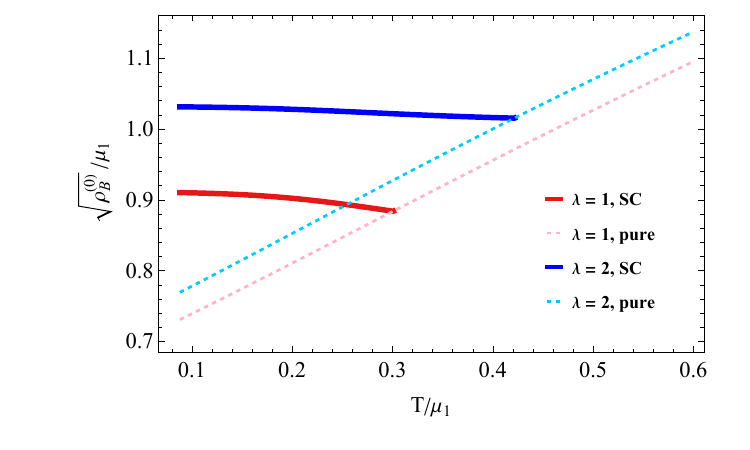} 
    \end{minipage}\hfill
    \begin{minipage}{0.51\textwidth}
        \centering
        \includegraphics[width=\textwidth]{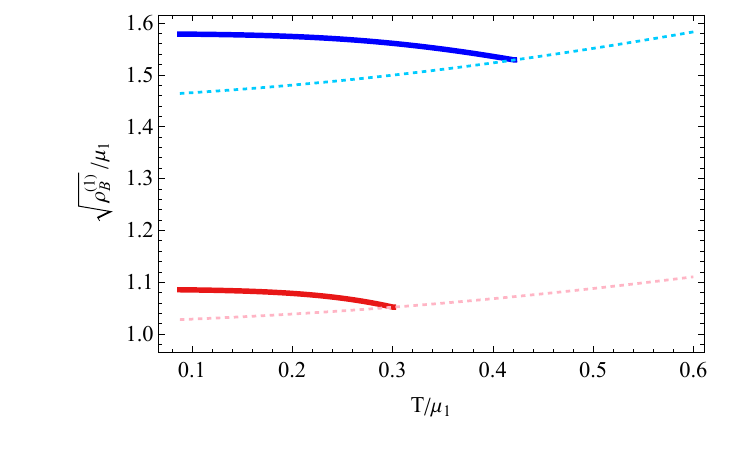} 
    \end{minipage}
    \vspace{0.5cm} 
    \begin{minipage}{0.49\textwidth}
        \centering
        \includegraphics[width=\textwidth]{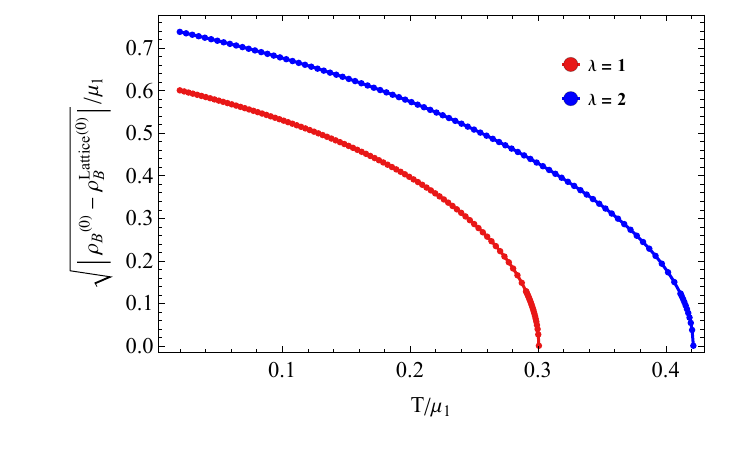} 
    \end{minipage}\hfill
    \begin{minipage}{0.51\textwidth}
        \centering
        \includegraphics[width=\textwidth]{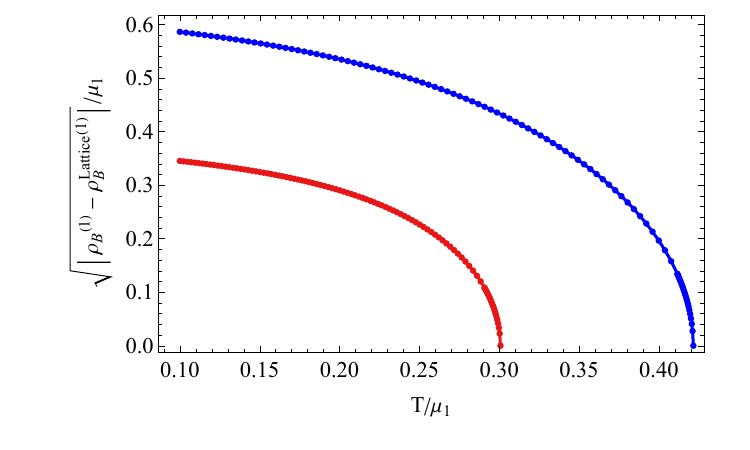} 
    \end{minipage}
    \caption{The zeroth and the first order of $\rho_B(x)$ and $\Delta \rho_B^{\text{SC}}(x)$ as the function of temperature T.}
    \label{fig:The zeroth to the SC third term of rho as the function of temperature T.}
\end{figure}

\begin{figure}[htp]
    \centering
    \begin{minipage}{0.5\textwidth}
        \centering
        \includegraphics[width=\textwidth]{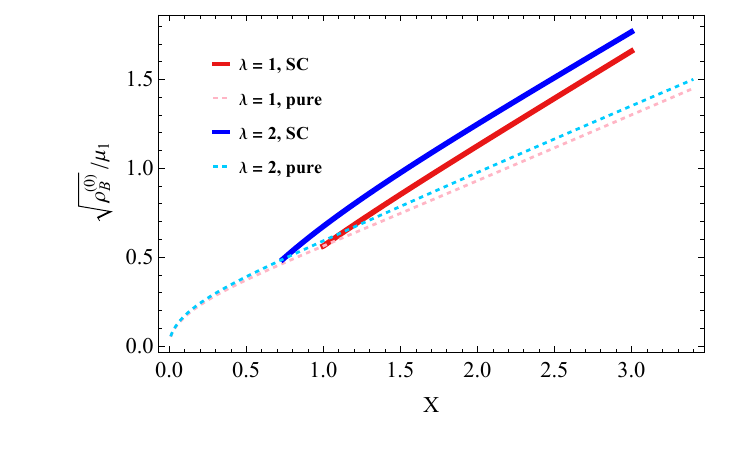} 
    \end{minipage}\hfill
    \begin{minipage}{0.5\textwidth}
        \centering
        \includegraphics[width=\textwidth]{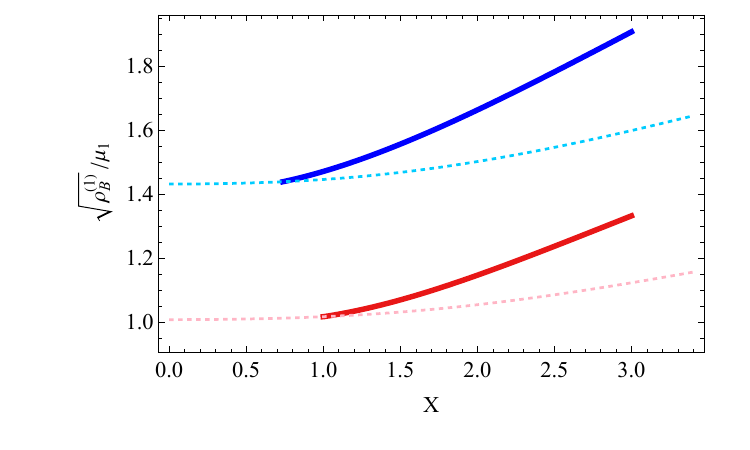} 
    \end{minipage}
    \vspace{0.5cm} 
    \begin{minipage}{0.5\textwidth}
        \centering
        \includegraphics[width=\textwidth]{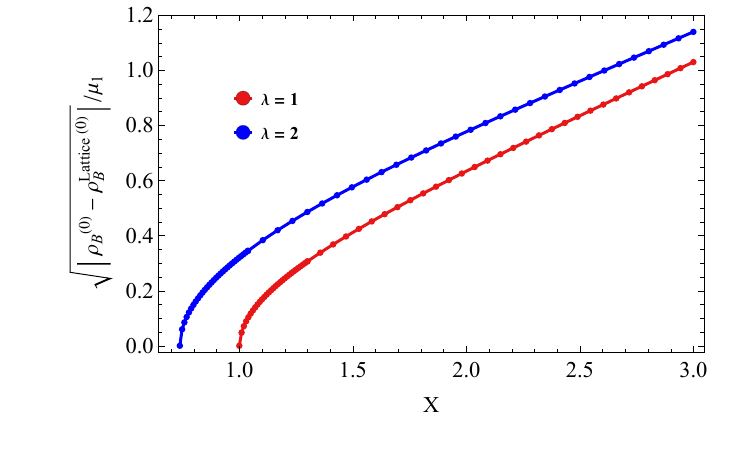} 
    \end{minipage}\hfill
    \begin{minipage}{0.5\textwidth}
        \centering
        \includegraphics[width=\textwidth]{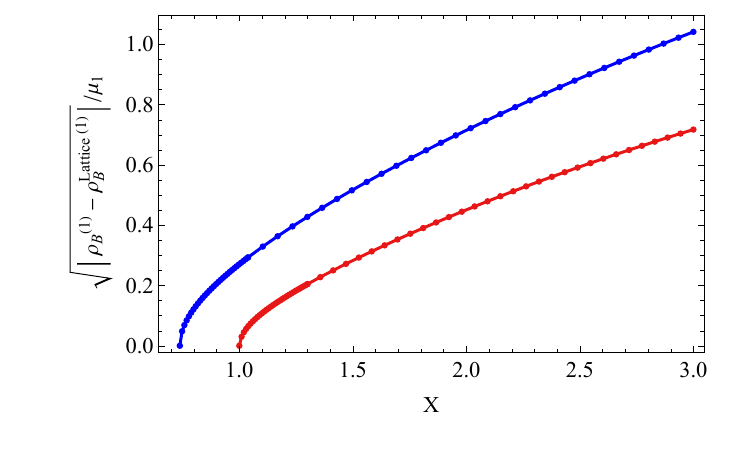} 
    \end{minipage}
    \caption{The zeroth and the first order of $\rho_B(x)$ and $\Delta \rho_B^{\text{SC}}(x)$ as the function of doping parameter X.}
    \label{fig:The zeroth to the third term of SC rho as the function of temperature X.}
\end{figure}
First, we illustrate the temperature dependence of the charge density $\rho_B$ and $\Delta \rho_B^{\text{SC}}$ in Fig. (\ref{fig:The zeroth to the SC third term of rho as the function of temperature T.}). In comparison with the charge density without condensation, we notice that all the orders of the charge density becomes larger after the phase transition. In particular, the discrepancy becomes more obvious when the temperature goes down.

This behavior is similar to the phenomenon observed in the CDW case. Notably, the condensation of superconductivity has a significant impact on the charge density, with changes in magnitude approximately 1000 times larger than those induced by the CDW phase (which are of order $10^{-3}$). Furthermore, for fixed $X$ and $T/\mu_1$, the leading orders of the charge density increase with the lattice amplitude $\lambda$. 

Next, we illustrate the doping dependence of the charge density $\rho_B$ and $\Delta \rho_B^{\text{SC}}$ in  Fig. (\ref{fig:The zeroth to the third term of SC rho as the function of temperature X.}). Similarly, we find that the leading orders of both $\rho_B(x)$ and $\Delta\rho_B^{\text{SC}}(x)$ increase with the increase of the doping parameter. In addition, the charge density becomes larger with increase of the lattice amplitude, indicating that the presence of the ionic lattice is helpful for the condensation of superconductivity.

\section{The Striped Superconducting Phase with ionic lattice}\label{sec5}
In this section we will investigate the interplay of CDW order and SC order by turning on both order parameters, with a focus on the effects of ionic lattice on the phase diagram.

\subsection{Phase diagram}
The striped superconducting phase is formed by the interplay of CDW order and SC order. Therefore, we turn on both scalar fields $\phi$ and $\eta$. In the absence of the ionic lattice, the phase diagram of this holographic model has previously been presented in \cite{Ling:2020qdd}, which illustrates the critical temperature of various phases as the function of the doping parameter $X$. In this paper, to observe the behavior of SSC under varying ionic lattice amplitudes $\lambda$, we plot the phase diagram on $T-X$ plane with a given lattice amplitude $\lambda$, which is illustrated in Fig. (\ref{fig:Phase diagram of CDW, SC and SSC}). Each diagram provides a detailed representation of the distribution of CDW, SC and SSC phases on $T-X$ plane. We remark that for these diagrams, we have fixed the wavenumber of the ionic lattice as $k/\mu_1=2$. As we demonstrated in the CDW section, from the beginning the CDW phase is locked at the commensurate state with $\tilde{p}/k=1$.

\begin{figure}[htp]
    \centering
    \begin{minipage}{0.45\textwidth}
        \centering
        \includegraphics[width=\textwidth]{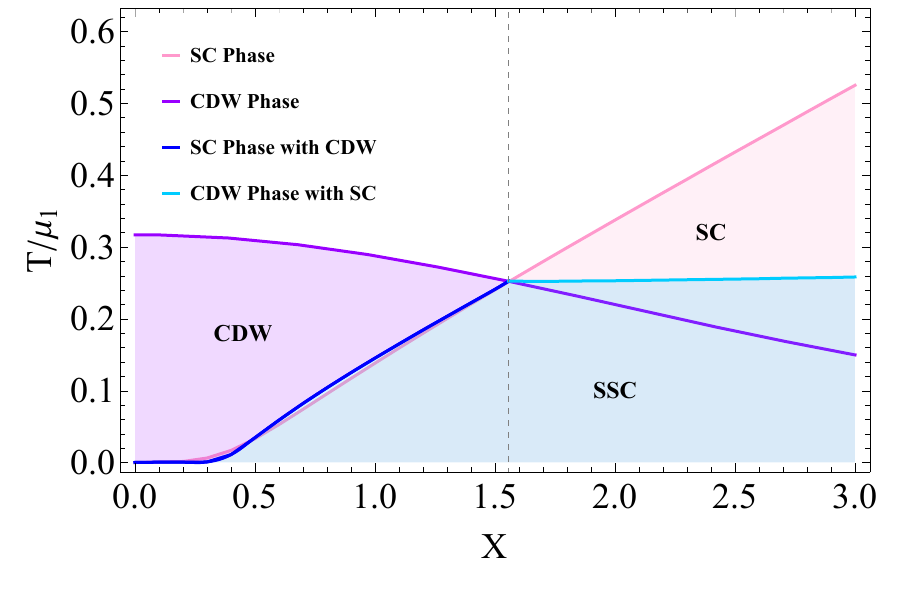}
    \end{minipage}
    \begin{minipage}{0.45\textwidth}
        \centering
        \includegraphics[width=\textwidth]{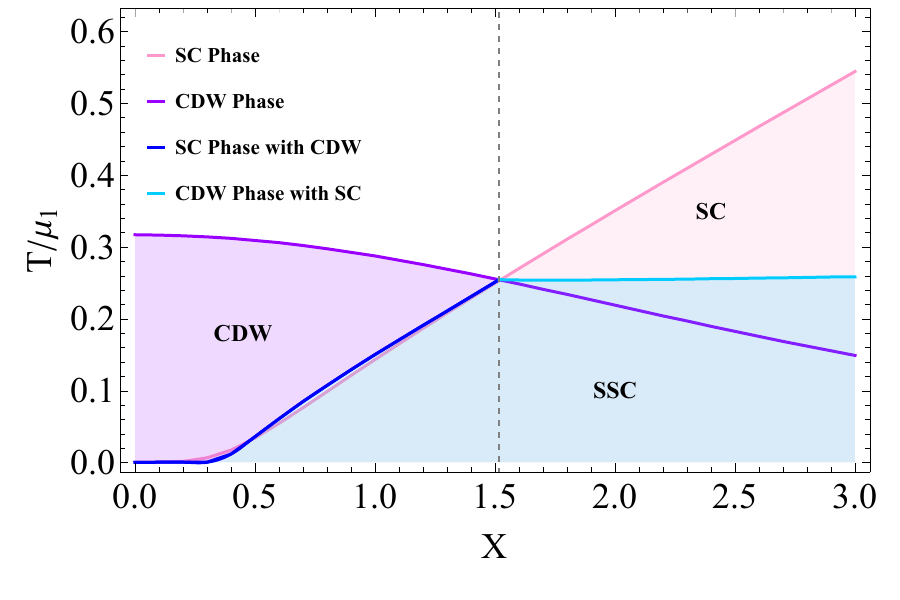}
    \end{minipage}
    \\
    \begin{minipage}{0.45\textwidth}
        \centering
        \includegraphics[width=\textwidth]{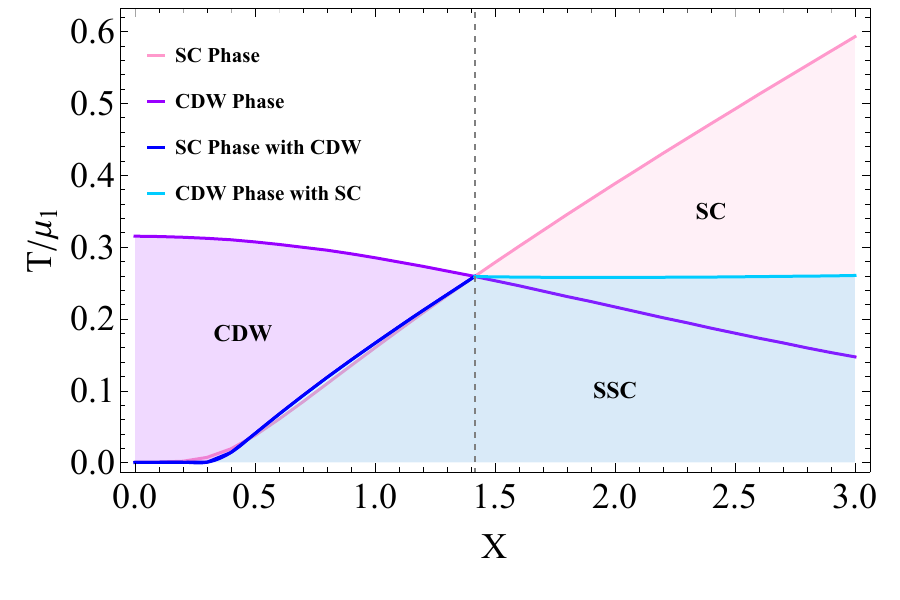}
    \end{minipage}
    \begin{minipage}{0.45\textwidth}
        \centering
        \includegraphics[width=\textwidth]{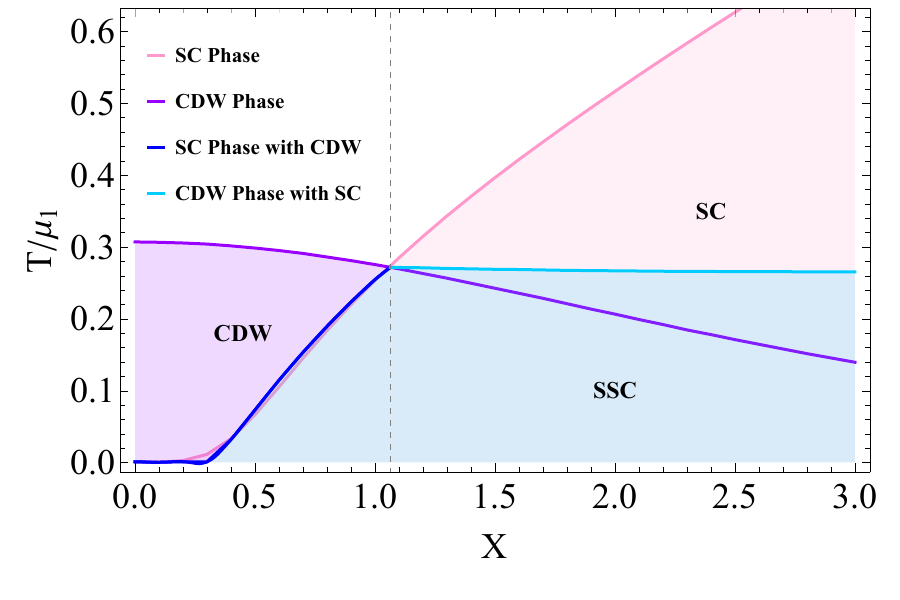}
    \end{minipage}
    \caption{Phase diagrams of CDW, SC, and SSC for different values of $\lambda$. The left panel on the top is for $\lambda=0$, while the right panel is for $\lambda=0.5$. At the bottom the left panel  is for $\lambda=1$, while the right panel is for $\lambda=2$.}
    \label{fig:Phase diagram of CDW, SC and SSC}
\end{figure}

The left panel on the top of Fig. (\ref{fig:Phase diagram of CDW, SC and SSC}) is the phase diagram for $\lambda=0$, which is almost the same as that one in \cite{Ling:2020qdd} (The discrepancy results from the different periodicity of the background with a wavenumber $k/\mu_1=2$). This diagram clearly reveals that the SSC phase emerges in the common region where both CDW and SC phases are allowed to coexist, reflecting that it is  the result of the interplay of CDW order and SC order. In the region with small doping parameter, the system enters a CDW phase at first with the dropping down of the temperature, followed by a transition to the SSC phase at lower temperature. Conversely, in the region with large doping parameter, the system enters a SC phase at first with the dropping down of the temperature, followed by a transition to the SSC phase at lower temperature. The above tendency gives rise to a critical doping parameter at $X_c\approx 1.55$. This diagram also reveals the competitive relation between the SC order and the CDW order at low temperature. In the region with small doping parameter, the presence of the CDW order does not change the critical temperature of SC much, as depicted by the deep blue curve in the diagram, while in the region with large doping parameter we notice the presence of the SC order evidently promote the critical temperature of CDW phase, as depicted by the blue curve which compares with the purple curve without SC in the diagram.

Now we turn to the phase diagram with an ionic lattice $\lambda=0.5$, which is illustrated as the right panel on the top of Fig. (\ref{fig:Phase diagram of CDW, SC and SSC}). First of all, we notice that the presence of the lattice does not change the structure of the phase diagram, which exhibits the similar behavior as the panel without lattice. Nevertheless, we observe a notable shift of the critical temperature for various phase transitions. Remarkably, the critical temperature of the SC phase experiences an obvious increase with the same doping parameter $X$, while the critical temperature  of CDW decreases a little bit. This suggests that the presence of the ionic lattice benefits the emergence of the SC phase, allowing it to persist over a broader temperature range.  As a result, we also notice that the critical point of the doping parameter shifts to the left at $X_c\approx 1.52$.

The phase diagrams for $\lambda=1$ and $\lambda=2$ are depicted as the left and right panel respectively at the bottom of Fig. (\ref{fig:Phase diagram of CDW, SC and SSC}), which further accentuates the enhanced stability of the SC phase with higher critical temperature. In contrast, the CDW phase exhibits a little bit lower critical temperature. The critical point of the doping parameter shifts to the left at $X_c\approx 1.41$ and $X_c\approx 1.06$ respectively. This trend underscores the growing influence of the ionic lattice in favoring superconductivity over charge density waves.

The variation of the phase diagram with the lattice amplitude collectively demonstrates the significant impact of ionic lattice on the delicate interplay between SC and CDW. As $\lambda$ increases, we observe a clear trend: the positive influence of the ionic lattice on SC growth becomes more pronounced, while simultaneously suppressing the CDW phase more effectively. This shift in phase stability and dominance provides valuable insights into the underlying mechanism governing these competing phenomena. Interestingly, the phase diagram observed in our model bears striking similarities to those found in some real superconductor systems. For instance, a comparable phase distribution is illustrated in \cite{PMID:38671053} for Kagome superconductors \cite{2023Natur.617..488Z,article121,article11,PMID:38671053,2024PhRvM...8a4801L}, suggesting that our findings may have broader implications for understanding complex phase behavior in various superconducting materials.

\subsection{The temperature behavior of order parameters $\phi$, $\eta$}\label{subsec:temp_behavior_order_params}

In this subsection, we examine the properties of the SSC phase, characterized by the coexistence of three orders, namely the pair density waves (PDW) order, CDW order and SC order, where the emergence of the PDW order is a consequence of the interplay between the CDW order and SC order.

First of all, we are concerned with the order parameter of PDW. In the absence of the ionic lattice, $\eta_2^{(1)}$ can be treated as the order parameter of PDW since it remains zero until the interplay between CDW and SC is involved, as performed in \cite{Ling:2020qdd}. However, due to  the presence of the ionic lattice, we point out that $\eta_2^{(1)}$ should not be treated as the order parameter of PDW any more since it is non-zero in SC phase without CDW. In this circumstance, we propose the quantity $\eta_2^{\text{PDW}}:=|\eta_2^{\text{SSC}}-\eta_2^{\text{SC}}|$ as the order parameter of PDW, where  $\eta_2^{\text{SSC}}$ is the solution of $\eta_2$ for SSC background containing both CDW and SC, while $\eta_2^{\text{SC}}$ is the solution of $\eta_2$ for SC lattice background without CDW at the same temperature. Specially, in the absence of the lattice, the subleading order of $\eta_2^{\text{PDW}}$, namely $\eta_2^{\text{PDW}(1)}$, goes back to the order parameter $\eta_2^{(1)}$. Similarly, the value of the PDW may be evaluated by $\Delta\rho_B^{\text{PDW}}(x):=(\rho_B^{\text{SSC}}(x)-\rho_B^{\text{SC}}(x)-\rho_B^{\text{CDW}}(x)+\rho_B^{\text{ionic}}(x))/\mu_1^2$), 
where $\rho_B^{\text{SSC}}(x)$, $\rho_B^{\text{SC}}(x)$, $\rho_B^{\text{CDW}}(x)$ and $\rho_B^{\text{ionic}}(x)$  are the charge densities over the SSC, SC (without CDW), CDW (without SC) and pure ionic (without SC and CDW) backgrounds, respectively.

\begin{figure}[htp]
    \centering
    \begin{minipage}{0.5\textwidth}
        \centering
        \includegraphics[width=\textwidth]{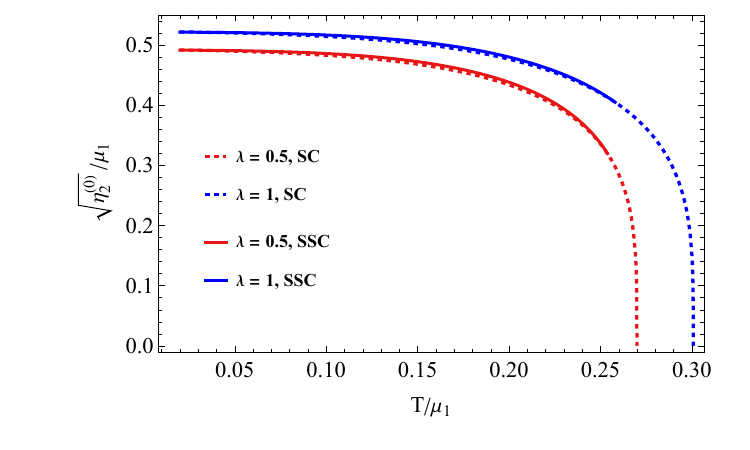} 
    \end{minipage}\hfill
    \begin{minipage}{0.5\textwidth}
        \centering
        \includegraphics[width=\textwidth]{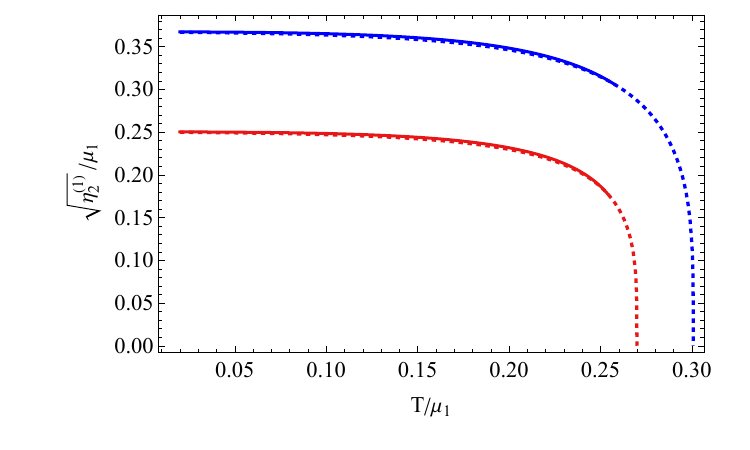} 
    \end{minipage}
    \vspace{0.5cm}
    \begin{minipage}{0.5\textwidth}
        \centering
        \includegraphics[width=\textwidth]{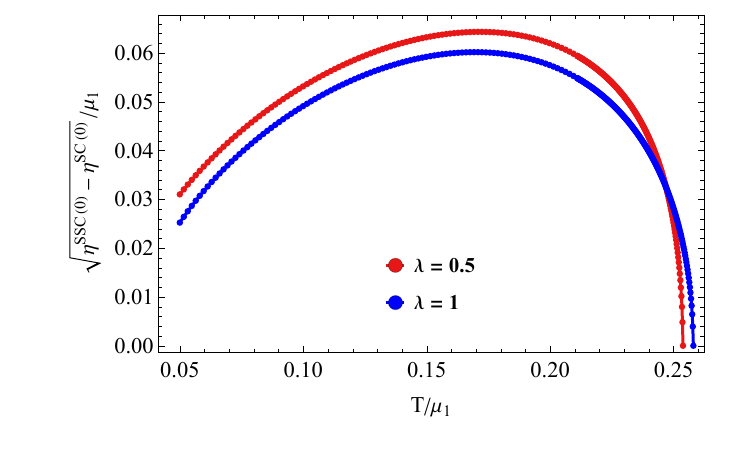} 
    \end{minipage}\hfill
    \begin{minipage}{0.5\textwidth}
        \centering
        \includegraphics[width=\textwidth]{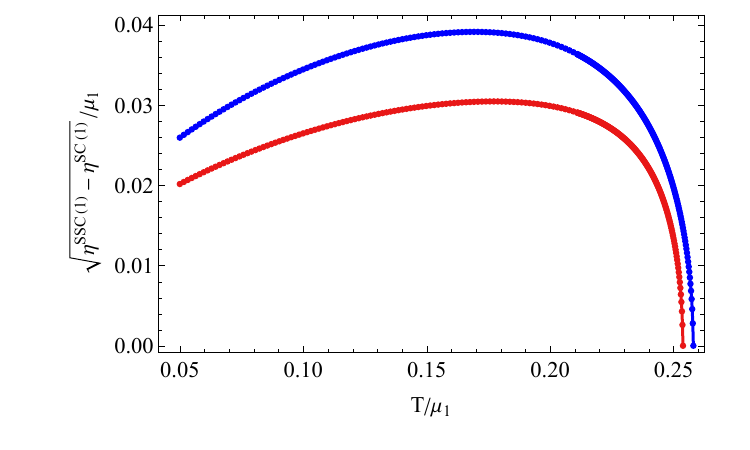} 
    \end{minipage}
    \caption{The SC order parameter $\eta^{(0)}_2$ and $\eta^{(1)}_2$ as the function of temperature T, where the doping parameter is fixed as $X=1.6$. The solid lines in color stand for $\eta^{(0)}_2$ and $\eta^{(1)}_2$ in SSC phase, while the dashed lines in color stand for $\eta^{(0)}_2$ and $\eta^{(1)}_2$ in the absence of CDW.}
    \label{fig:image23}
\end{figure}

\begin{figure}[htp]
    \centering
    
    \begin{minipage}{0.5\textwidth}
        \centering
        \includegraphics[width=\linewidth]{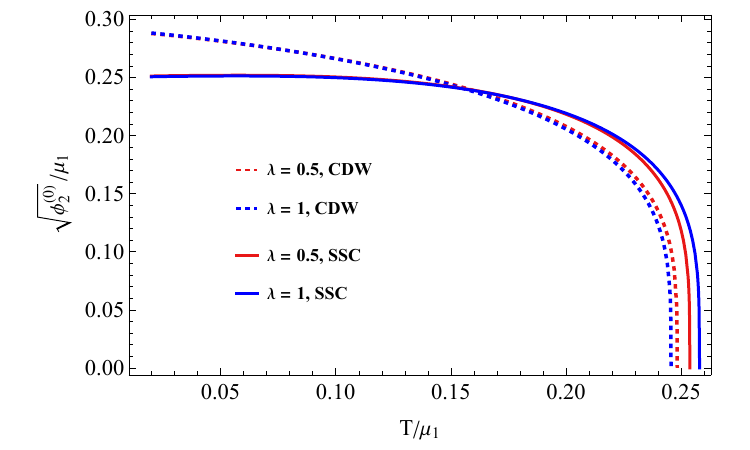} 
    \end{minipage}\hfill 
    \begin{minipage}{0.5\textwidth}
        \centering
        \includegraphics[width=\linewidth]{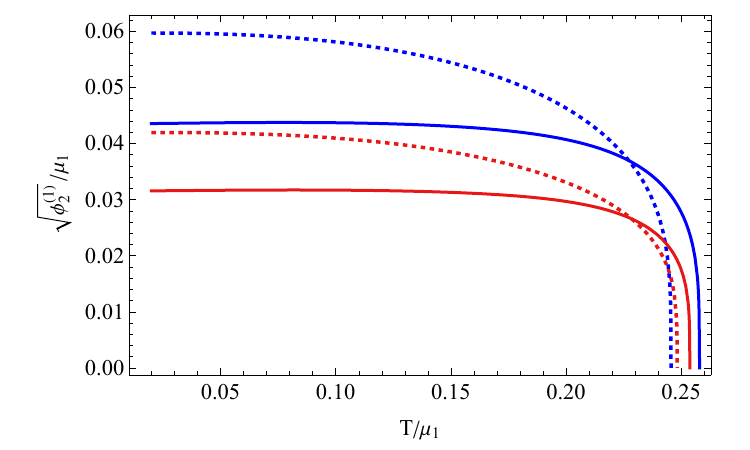} 
    \end{minipage}
    \caption{
    The CDW order parameter $\phi^{(0)}_2$ and $\phi^{(1)}_2$ as the function of temperature T, where the doping parameter is fixed as $X=1.6$. The solid lines in color stand for $\phi^{(0)}_2$ and $\phi^{(1)}_2$ in SSC phase, while the dashed lines in color stand for $\phi^{(0)}_2$ and $\phi^{(1)}_2$ in the absence of SC.
    }
    \label{fig:image2}
\end{figure}

Fig. (\ref{fig:image23}) illustrates the temperature dependence of the first two leading orders of $\eta_2$, namely $\eta_2^{(0)}$ and $\eta_2^{(1)}$, where  dashed lines represent values in the SC phase without CDW, while solid lines represent values in the SSC phase. Then the contribution of PDW may be evaluated by the discrepancy of these two order parameters, namely  $\eta_2^{\text{PDW}}$. For simplicity and without the loss of generality, in the following subsections we only demonstrate the data with $\lambda=0.5$ and $1$. As demonstrated in the left panel of Fig. (\ref{fig:image23}), we notice that  $\eta_2^{\text{PDW}}$ does not exhibit a monotonous behavior with the dropping down of the temperature, which could increase at first and then decrease gradually. In addition, we find  $\eta_2^{\text{PDW}(1)}$  becomes larger with the increase of the lattice amplitude, indicating that the presence of lattice will enhance the formation of PDW.

In the end of this subsection we briefly demonstrate the temperature dependence of the CDW order parameters, namely $\phi_2^{(0)}$ and $\phi_2^{(1)}$ in Fig. (\ref{fig:image2}), where solid lines represent the value of order parameters  in the SSC phase, while dashed lines represent the values of order parameters in the CDW phase without SC. It is noticed that both order parameters become saturated with the dropping down of the temperature. In particular, the saturation value in the CDW phase is a little bit higher than that in the SSC phase with the same lattice amplitude. This phenomenon suggests that superconductivity suppresses the formation of CDW. It could be understood as the result of the competition between the CDW order and SC order such that some carriers in the CDW are borrowed to form the PDW. In addition, we find the saturated value of the leading order $\phi_2^{(0)}$ does not change much with the variation of the lattice amplitude but the saturated value of the subleading order $\phi_2^{(1)}$ increases with the lattice amplitude, which is similar to the phenomenon in CDW phase.

\subsection{The temperature behavior of charge density $\rho_B$}\label{subsec:temp_behavior_charge_density}
In this subsection we present the temperature dependence of charge density $\rho_B$ in SSC phase. 

\begin{figure}[!h]
    \centering
    \begin{minipage}{0.5\textwidth}
        \centering
        \includegraphics[width=\textwidth]{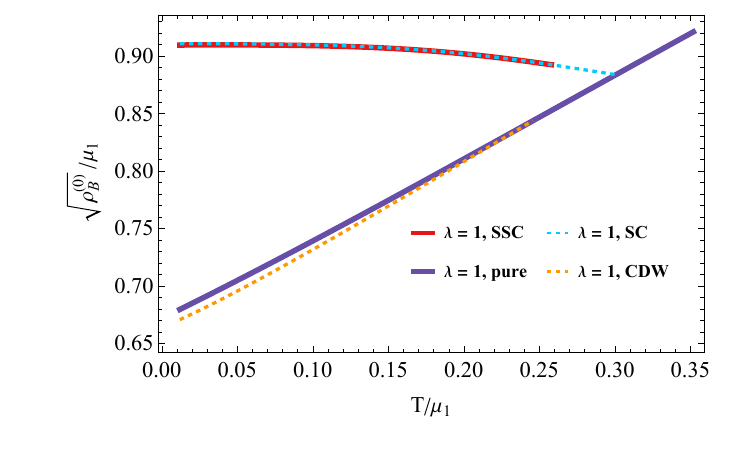} 
    \end{minipage}\hfill
    \begin{minipage}{0.5\textwidth}
        \centering
        \includegraphics[width=\textwidth]{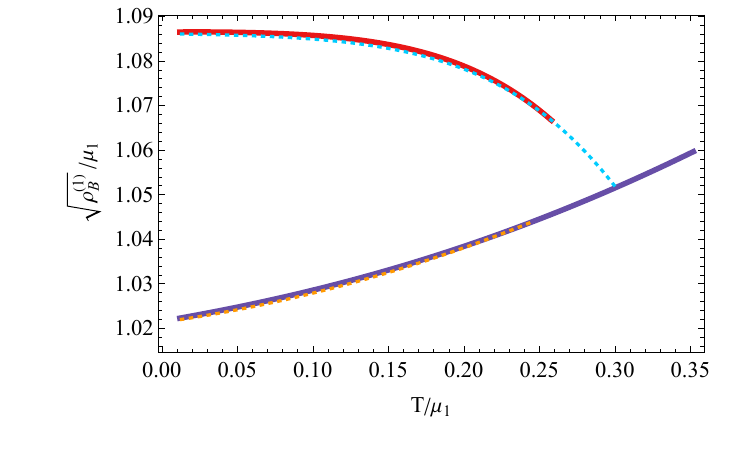} 
    \end{minipage}
    \vspace{0.5cm} 
    \begin{minipage}{0.5\textwidth}
        \centering
        \includegraphics[width=\textwidth]{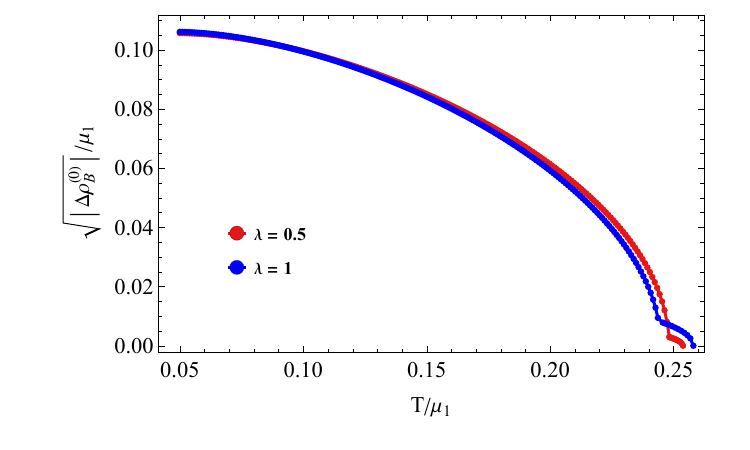} 
    \end{minipage}\hfill
    \begin{minipage}{0.5\textwidth}
        \centering
        \includegraphics[width=\textwidth]{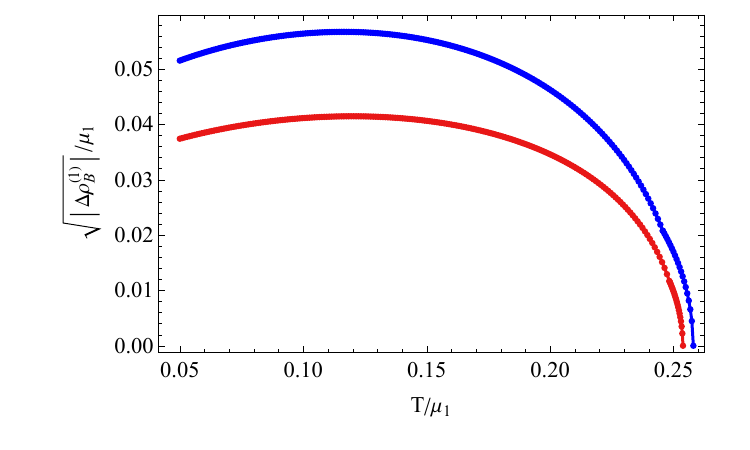} 
    \end{minipage}
    \caption{The charge density $\rho_B ^{(0)}$ and $\rho^{(1)}_B$ as the function of temperature T, where the doping parameter is fixed as $X=1.6$. Two figures on the top show the total charge density $\sqrt{\rho_B ^{(0)}}/\mu_1$ and $\sqrt{\rho_B ^{(1)}}/\mu_1$, while two figures at the bottom show the total difference among SSC and CDW, SC phase, which is defined as the charge density of PDW $\Delta\rho_B^{\text{PDW}}$ as before.}\label{fig:image25}
\end{figure}

Firstly, two figures on the top of Fig. (\ref{fig:image25}) show the first two leading orders of total charge density $\rho_B$ across all phases. As expected, the behavior of $\rho_B$ in the CDW, SC, and pure ionic phases remains consistent with the results discussed in previous sections. Notably, the charge density $\rho_B$ in the SSC phase closely resembles that of the SC phase, highlighting their similarity.

Secondly, two figures at the bottom of Fig. (\ref{fig:image25}) depict the first two leading orders of  $\Delta\rho_B^{\text{PDW}}$, which is defined in the previous section to measure the distribution of PDW.
The figures show that the leading order increases steadily as the temperature decreases, while the subleading order grows initially, reaches a maximum, and then gradually decreases.

\newpage
\subsection{The doping behavior of the order parameter $\eta$}\label{subsec:doping_behavior_order_params}
\begin{figure}[t]
    \centering
    \begin{minipage}{0.5\textwidth}
        \centering
        \includegraphics[width=\textwidth]{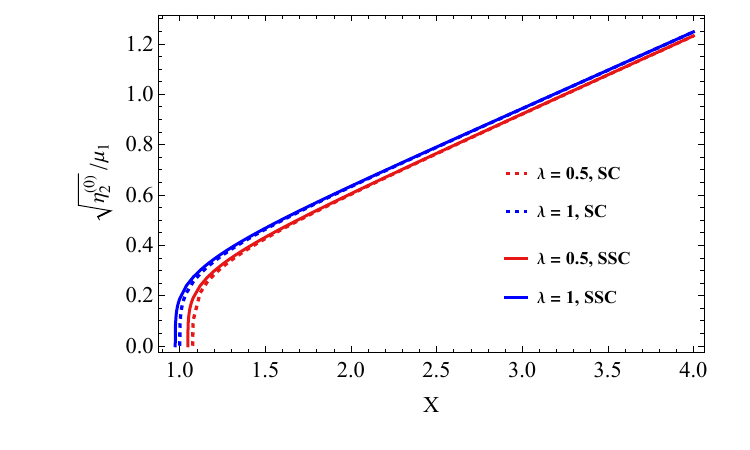} 
    \end{minipage}\hfill
    \begin{minipage}{0.5\textwidth}
        \centering
        \includegraphics[width=\textwidth]{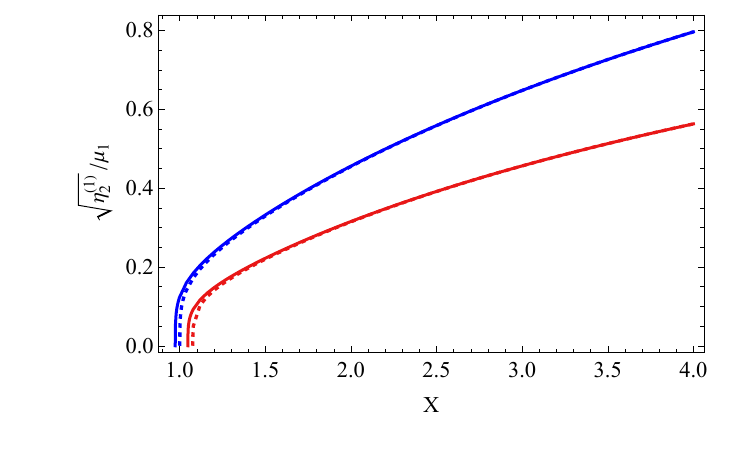} 
    \end{minipage}
    \vspace{0.5cm} 
    \begin{minipage}{0.5\textwidth}
        \centering
        \includegraphics[width=\textwidth]{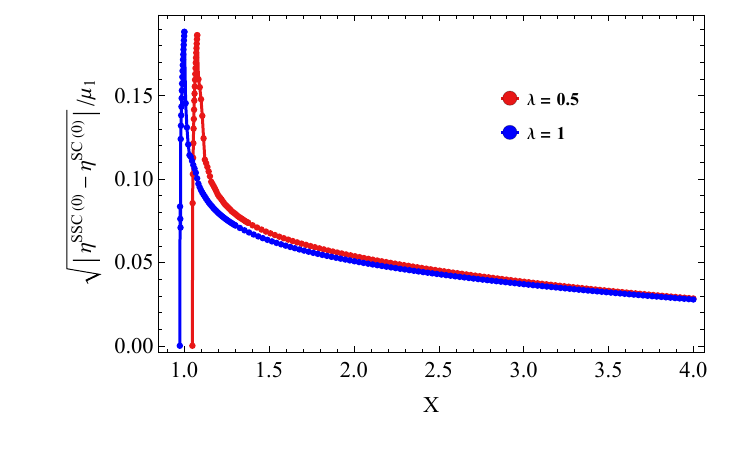} 
    \end{minipage}\hfill
    \begin{minipage}{0.5\textwidth}
        \centering
        \includegraphics[width=\textwidth]{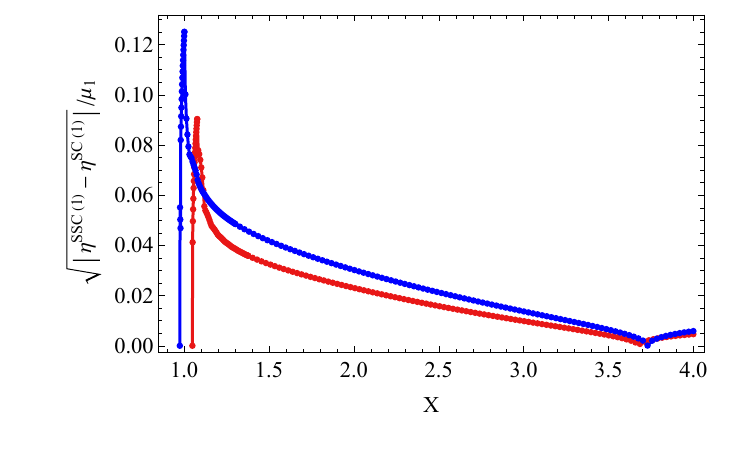} 
    \end{minipage}
    \caption{The SC order parameter $\eta^{(0)}_2$ and $\eta^{(1)}_2$ as the function of doping parameter X, where the temperature is fixed as $T/\mu_1=0.16$. The solid lines in color stand for $\eta^{(0)}_2$ and $\eta^{(1)}_2$ in SSC phase, while the dashed lines in color stand for $\eta^{(0)}_2$ and $\eta^{(1)}_2$ in the absence of CDW.}
    \label{fig:image24}
\end{figure}

In this subsection we show the doping dependence of the order parameter $\eta$ in SSC phase. Two figures on the top of Fig. (\ref{fig:image24}) demonstrate that the condensation value increases with the doping parameter in both SC and SSC phases. Interestingly, we also observe that the zeroth-order coefficient $\eta_2^{(0)}$ increases with $\lambda$, but the first-order coefficient $\eta_2^{(1)}$ increases more significantly with $\lambda$, particularly in the region with larger doping parameter. 
In contrast, two figures at the bottom reveal that the PDW order, represented by ($\eta_2^{\text{SSC}} - \eta_2^{\text{SC}}$), initially reaches a maximum at a small doping level and subsequently decreases as doping increases. 

\subsection{The doping behavior of charge density $\rho_B$}\label{subsec:temp_behavior_charge_density}
In this subsection we present the doping dependence of charge density $\rho_B$ in the SSC phase at a fixed temperature.  

\begin{figure}[htp]
    \centering
    \begin{minipage}{0.5\textwidth}
        \centering
        \includegraphics[width=\textwidth]{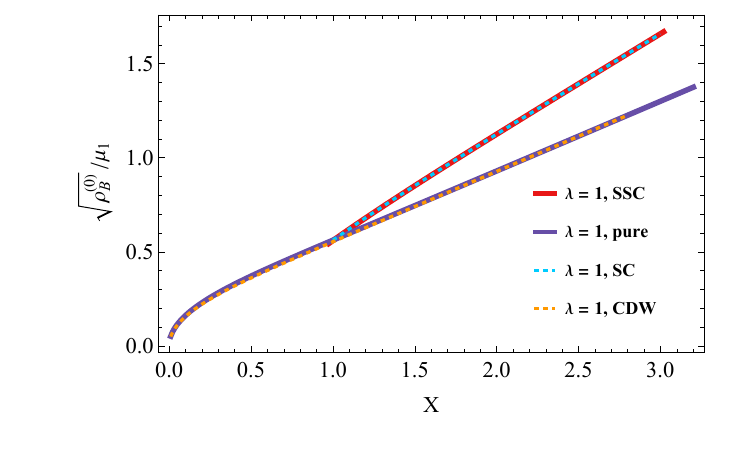} 
    \end{minipage}\hfill
    \begin{minipage}{0.5\textwidth}
        \centering
        \includegraphics[width=\textwidth]{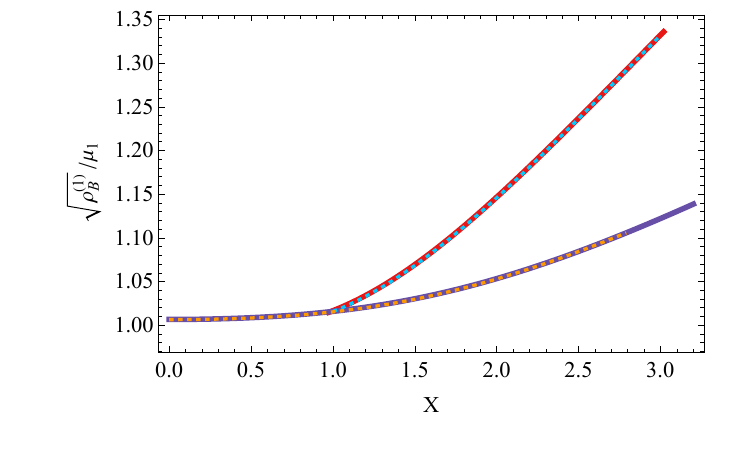} 
    \end{minipage}
    \vspace{0.5cm} 
    \begin{minipage}{0.5\textwidth}
        \centering
        \includegraphics[width=\textwidth]{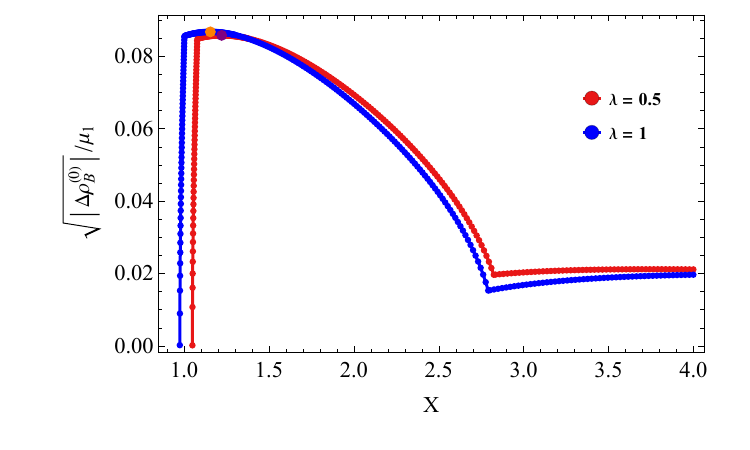} 
    \end{minipage}\hfill
    \begin{minipage}{0.5\textwidth}
        \centering
        \includegraphics[width=\textwidth]{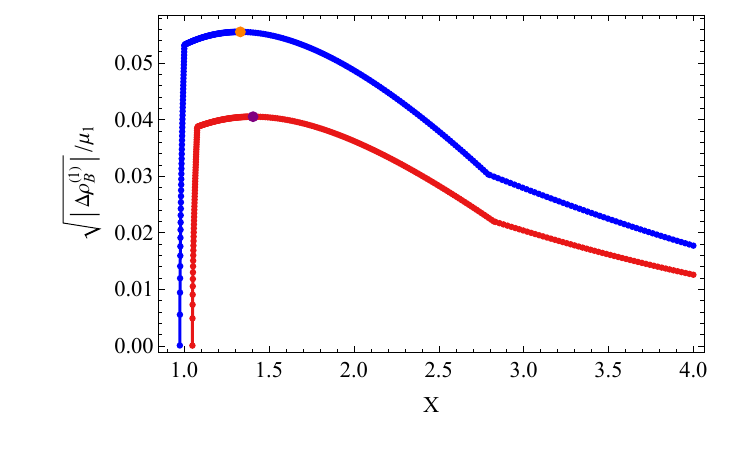} 
    \end{minipage}
    \vspace{0.5cm} 
    \begin{minipage}{0.5\textwidth}
        \centering
        \includegraphics[width=\textwidth]{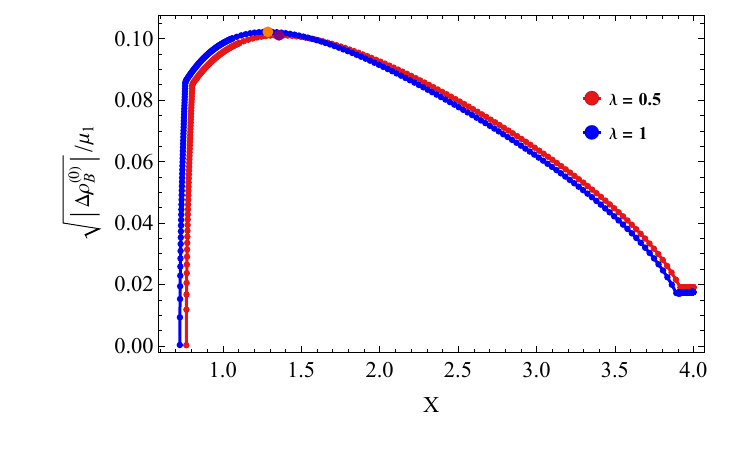} 
    \end{minipage}\hfill
    \begin{minipage}{0.5\textwidth}
        \centering
        \includegraphics[width=\textwidth]{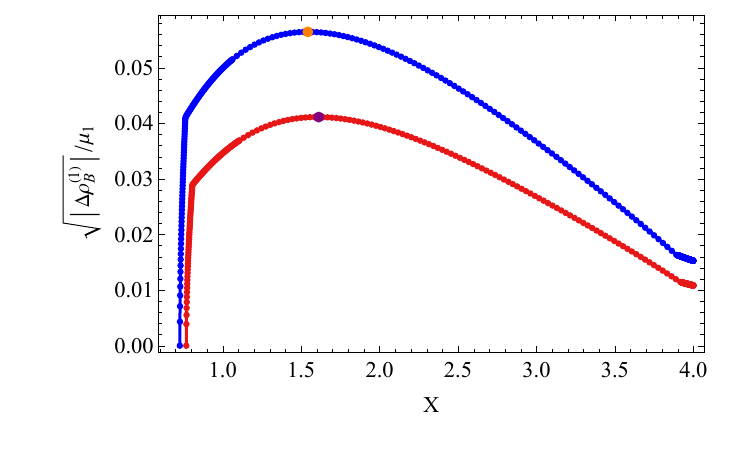} 
    \end{minipage}
    \caption{The charge density $\rho_B ^{(0)}$ and $\rho^{(1)}_B$ as the function of doping parameter X with fixed temperature. The top two figures show the total charge density $\sqrt{\rho_B ^{(0)}}/\mu_1$ and $\sqrt{\rho_B ^{(1)}}/\mu_1$ with $T/\mu_1=0.16$. The middle two show the charge density of PDW $\Delta\rho^{(0)}_B$ and $\Delta\rho^{(1)}_B$ with $T/\mu_1=0.16$. The bottom two show the charge density of PDW $\Delta\rho^{(0)}_B$ and $\Delta\rho^{(1)}_B$ with $T/\mu_1=0.1$.}\label{fig:image26}
\end{figure}

Firstly, we set $T/\mu_1=0.16$. Two figures on the top of Fig. (\ref{fig:image26}) display the total charge density across the CDW, SC, SSC, and pure ionic phases during the course of varying the doping parameter $X$. Consistent with previous results, the charge density in the SSC phase closely resembles that in the SC phase. Additionally, the charge density in all phases increases with the doping parameter. Interestingly, the increase in the ionic amplitude $\lambda$ has a more pronounced effect on the first-order coefficient $\rho_B^{(1)}$ and higher-order terms, while its influence on the zeroth-order coefficient $\rho_B^{(0)}$ is relatively minor.

Secondly, two figures in the middle of Fig. (\ref{fig:image26}) depict the charge density associated with the PDW component. Notably, the PDW charge density reaches a maximum at a doping parameter different from the one where $\eta^{\text{PDW}}$ reached its peak in the previous subsection. In the figures, these distinct maxima are marked by orange and purple points. This result highlights the competitive interplay between the CDW and SC orders, demonstrating that PDW order can achieve its peak value at specific doping levels due to the interaction between these two orders. Also it mirrors the qualitative behavior of the pair order parameter's response to doping variations in condensed matter physics \cite{2024Sci...384h7691X}.

Moreover, we present an additional result at a lower temperature $T/\mu_1=0.1$ as shown in the last two pictures in Fig. (\ref{fig:image26}). At this lower temperature, the existence of a maximum point becomes more evident, and the corresponding doping parameter at which it occurs shifts to a higher value. Interestingly enough, we notice that at lower temperature the optimal doping parameter is closer to the critical doping parameter which is about $X_c\approx 1.52$ for $\lambda=0.5$ and $X_c\approx 1.41$ for $\lambda=1.0$. In this doping region both the CDW and SC have visible contributions to the system, leading to an optimal doping for the production of PDW. Therefore, the statement that the PDW component is the consequence of the interplay between the CDW order and SC  order is justified.

\subsection{Free energy of all phases}

From the phase diagram, we observe that as the temperature decreases, the system undergoes two distinct pathways to the SSC phase, depending on the doping parameter $X$. For $X < X_c$, where $X_c$ is the critical doping, the system transitions from a metallic phase to a CDW phase, followed by a transition to the SSC phase. In contrast, for $X > X_c$, the system first undergoes a transition from the metallic phase to the SC phase, 
and then transitions to the SSC phase.

\begin{figure}[htp]
    \centering
    \includegraphics[width=0.8\linewidth]{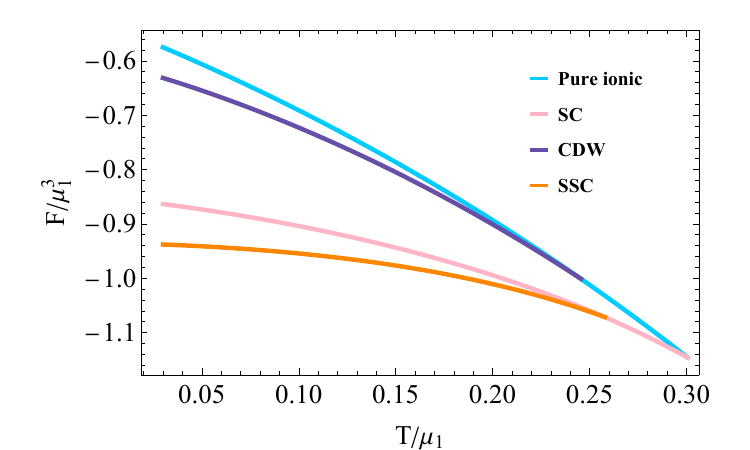} 
    \caption{The averaged free energy for RN black hole, CDW/SC black hole and SSC black hole with the same ionic lattice, where $X = 1.6$ and $\lambda = 1$.}
    \label{fig:freeenergy3phases}
\end{figure}

Without loss of generality, we compute the free energy for the case of $X > X_c$ with $\lambda=1$. The results, presented in Fig. (\ref{fig:freeenergy3phases}), demonstrate that the SSC phase consistently exhibits the lowest average free energy compared to the SC and CDW phases. This indicates that, in the presence of an ionic lattice, the SSC phase is indeed the most stable phase among these competing phases.

\section{Conclusion and Discussion}\label{sec6}
In this paper we have investigated the striped superconductor in a holographic model with ionic lattices. This model is featured by  a phase diagram with three distinct phases, namely the CDW phase, ordinary SC phase and the SSC phase. In particular, the formation of PDW in SSC phase is ascribed to the interplay between the CDW order and the ordinary SC order. In CDW phase, we have manifestly demonstrated that  due to the periodic nature of the lattice background, different types of CDW solutions can be found at the same temperature below the critical temperature, which justifies the cartoon phase diagram which was originally sketched in \cite{Andrade:2017leb}. More importantly, with the increase of the lattice amplitude these solutions are locked in different commensurate states. Nevertheless, we have shown that Type I solutions are always the most stable solutions since they exhibit the lower free energy. Thus with a focus on Type I solutions, we have demonstrated the feature of the order parameter and the charge density with the variation of the temperature $T$, the doping parameter $X$ as well as the lattice amplitude $\lambda$.  It is found  that with the increase of the lattice amplitude, the critical temperature of CDW phase goes down and the density of CDW becomes smaller, both effects indicating that the presence of the ionic lattice suppresses the formation of CDW. On the contrary, in SC phase it is found that with the increase of the lattice amplitude, the critical temperature of SC phase goes up and the charge density   becomes larger, both effects indicating that the presence of the ionic lattice benefits the formation of SC. In SSC phase, which is characterized by the coexistence of the CDW order and SC order, we have proposed a notion to describe the contribution of PDW, which is described by the discrepancy of the charge density in SSC phase and SC phase. It is shown that the density of PDW increases with the dropping down of the temperature. Remarkably, we have demonstrated there exists an optimum doping for the formation of PDW in this holographic model, which is qualitatively similar to the graphical representation of the pair order parameter with the variation of the doping in condensed matter physics \cite{2024Sci...384h7691X}.  Furthermore, we have verified that the SSC phase has the lowest free energy among all three phases, thus it is the favorite state that the material intends to stay as both CDW and SC orders are involved.

In comparison with all the previous work in holographic literature, we have made the following substantial progress on the road towards understanding on the nature of high-temperature superconductivity by holography. First of all, most of the essential ingredients needed to describe the phase diagram of high-temperature superconductivity have been taken into account in our current work. By virtue of the ionic lattices,  the notion of commensurate states has been introduced and the lock-in effect has been observed for CDW phase, which plays an essential role in constructing the Mott insulator by holography, as revealed in \cite{Andrade:2017ghg}. In this paper we have pushed this approach forward by adding the superconducting phase into the setup and the behavior of the PDW in commensurate state with $\tilde{p}/k=1$ has been revealed. By virtue of two-gauge-field formalism, the notion of doping has been introduced and the dependent behavior of the phase transition on the doping can be explicitly observed. Furthermore, by introducing the order parameters for CDW and SC separately, one is able to implement the CDW phase and SC phase by different phase transitions such that the picture of PDW formed by the interplay between CDW order and SC order becomes vivid.  

We leave the following open questions for future investigation. Firstly, the transport properties of the model could be explored by computing the optical conductivity using the linear response method. Secondly, simulating the phase diagram of high-temperature superconductivity remains a significant challenge. Incorporating additional interactions among the various orders could make the phase diagram more realistic. Lastly, further study of different types of CDW solutions could provide deeper insights into their relationship with SC and PDW orders.
\section*{Acknowledgments}
We are very grateful to Pan Li, Wen-bin Pan  and Zhangping Yu for helpful discussions.  This work is supported in part by the Natural Science Foundation of China (Grant Nos.~12035016,~12275275,~12475054). It is also supported by Beijing Natural Science Foundation (Grant No.~1222031) and the Innovative Projects of Science and Technology (E2545BU210) at IHEP.

\appendix
\section{Holographic renormalization of the bulk action}\label{appendix:A}
The total action \cite{1999CMaPh.208..413B,Donos:2013wia,Cremonini:2017usb} is given by
\begin{equation}
    S_{\text{total}}=S_{\text{bulk}}+\frac{1}{\kappa^2}\int_{\partial \mathcal{M}}\sqrt{-\gamma}\left(K-\frac{2}{l}\right).
\end{equation}
By taking the variation of  the action we get
\begin{equation}
    T_{\mu\nu}=\frac{1}{\kappa^2}\left(K\gamma_{\mu\nu}-K_{\mu\nu}-\frac{2}{l}\gamma_{\mu\nu}\right),\red{\label{emt}}
\end{equation}
where $K_{\mu\nu}=\frac{1}{2}\left(\nabla_\mu n_\nu+\nabla_\nu n_\mu\right)|_{z\rightarrow 0}$, $K=g^{\mu\nu}\nabla_\mu n_\nu |_{z\rightarrow 0}$, and $n^a=\left. z\sqrt{\frac{(1-z)p(z)}{Qzz(x,z)}}\left(\dfrac{\partial}{\partial z}\right)^a \right|_{z\rightarrow 0}$ is the outward pointing normal vector to the boundary $\partial \mathcal{M}$. In addition, $n_a=g_{ab}n^b|_{z\rightarrow 0}$ and $\gamma_{\mu\nu}=g_{\mu\nu}-n_\mu n_\nu$ is the induced metric near the boundary with $z\rightarrow 0$.

Inserting the asymptotic expansion of the metric  in (\ref{eq:7}) into (\ref{emt}) and setting the AdS radius $l=\frac{1}{2}$, we get
\begin{equation}
    \begin{aligned}
    T_{tt}&=\frac{1}{2\kappa^2}(16+\mu_1^2+\mu_2^2+24 q_{xx}(x)+24 q_{yy}(x))z+\mathcal{O}(z^2),\\
    T_{xx}&=\frac{1}{16\kappa^2}(16+\mu_1^2+\mu_2^2-48 q_{tt}(x)-48 q_{yy}(x))z+\mathcal{O}(z^2),\\
    T_{yy}&=\frac{1}{16\kappa^2}(16+\mu_1^2+\mu_2^2-48 q_{tt}(x)-48 q_{xx}(x))z+\mathcal{O}(z^2).
\end{aligned}
\end{equation}

We can rewrite the induced metric on the boundary in ADM form as 
\begin{equation}
    ds^2=-N_{\Sigma}^2 dt^2+h_{\mu\nu}dx^\mu dx^\nu.
\end{equation}
Then the mass $M$ is
\begin{equation}
    M=\int dx^2 \sqrt{h}N_{\Sigma}u^t u^t T_{tt}=\int dx^2 \frac{1}{2z} T_{tt},
\end{equation}
where in our case $\sqrt{h}=\dfrac{1}{z^2}$, $N_{\Sigma}=\dfrac{2}{z}$ and $u^t=\dfrac{z}{2}$ is the timelike unitary vector on the boundary.

Thus
\begin{equation}
    M=\int dx^2 \frac{1}{4\kappa^2}(16+\mu_1^2+\mu_2^2+24 q_{xx}(x)+24 q_{yy}(x)).
\end{equation}

The mass density $m(x)$ is then read as 
\begin{equation}
    m(x)=\frac{1}{4\kappa^2}(16+\mu_1^2+\mu_2^2+24 q_{xx}(x)+24 q_{yy}(x)).
\end{equation}

Similarly, the momentum $P$ is 
\begin{equation}
\begin{aligned}
        P_x&=\int dx^2 \sqrt{h}h_{x}^{\ i} u^x T_{i x}=\int dx^2 \frac{1}{z} T_{xx},\\
        P_y&=\int dx^2 \sqrt{h}h_{y}^{\ i} u^y T_{i y}=\int dx^2 \frac{1}{z} T_{yy}.
\end{aligned}
\end{equation}
where $h_{i}^{\ j}=\delta_{i}^{\ j}$ and  $u^i=z$ is the spacelike unitary vector on the boundary.

Therefore we obtain the energy-momentum tensor on the boundary
\begin{equation}
    \begin{aligned}
    \tilde{T}_{tt}&=\frac{1}{4\kappa^2}(16+\mu_1^2+\mu_2^2+24 q_{xx}(x)+24 q_{yy}(x)),\\
    \tilde{T}_{xx}&=\frac{1}{16\kappa^2}(16+\mu_1^2+\mu_2^2-48 q_{tt}(x)-48 q_{yy}(x)),\\
    \tilde{T}_{yy}&=\frac{1}{16\kappa^2}(16+\mu_1^2+\mu_2^2-48 q_{tt}(x)-48 q_{xx}(x)).
\end{aligned}
\end{equation}
The metric of the boundary theory we are interested in is 
\begin{equation}
    \gamma_{\text{boundary}}=-4 dt^2 +dx^2 +dy^2.
\end{equation}
Using the conditions in (\ref{eq:qttqxx}), it is easy to find that $\tilde{T}^\mu_{\ \mu}=0$ and $\partial^\mu \tilde{T}_{\mu\nu}=0$.

\bibliographystyle{style1}
\bibliography{holographic_SSC}

\end{document}